\documentclass[10pt,twocolumn,superscriptaddress,aps,pra,balancelastpage]{revtex4-2}
\usepackage[utf8]{inputenc}
\usepackage{amsmath,amsfonts,amssymb,fullpage,graphicx,physics,braket}
\usepackage[dvipsnames]{color,xcolor,colortbl}
\usepackage{hyperref}
\hypersetup{
  colorlinks=true,
  linkcolor=NavyBlue
}

\usepackage{CJKutf8}

\newcommand{\xiCorr}{{\xi}}

\begin{document}

\title{Exponential entanglement advantage in sensing correlated noise}

\author{\begin{CJK}{UTF8}{gbsn}Yu-Xin Wang (王语馨)\end{CJK}}
\email{yxwang.physics@outlook.com}
\affiliation{Joint Center for Quantum Information and Computer Science, NIST/University of Maryland, College Park, MD, 20742, USA}

\author{{Jacob Bringewatt}}
\affiliation{Joint Center for Quantum Information and Computer Science, NIST/University of Maryland, College Park, MD, 20742, USA}
\affiliation{Department of Physics, Harvard University, Cambridge, MA 02138 USA}
\affiliation{Joint Quantum Institute, NIST/University of Maryland, College Park, MD, 20742, USA}

\author{Alireza Seif}
\affiliation{IBM Quantum, IBM T.J. Watson Research Center, Yorktown Heights, NY 10598, USA}

\author{Anthony J. Brady}
\affiliation{Joint Center for Quantum Information and Computer Science, NIST/University of Maryland, College Park, MD, 20742, USA}
\affiliation{Joint Quantum Institute, NIST/University of Maryland, College Park, MD, 20742, USA}

\author{Changhun Oh}
\affiliation{Department of Physics, Korea Advanced Institute of Science and Technology, Daejeon 34141, Korea}

\author{{Alexey~V.~Gorshkov}}
  \affiliation{Joint Center for Quantum Information and Computer Science, NIST/University of Maryland, College Park, MD, 20742, USA}
  \affiliation{Joint Quantum Institute, NIST/University of Maryland, College Park, MD, 20742, USA}

\begin{abstract} 
Correlated noise naturally arises in quantum systems due to, e.g., phase and amplitude fluctuations in global control or long-range correlations generated by a many-body sensing target near criticality.
In this Letter, we propose a new form of exponential quantum advantage in the context of sensing correlated noise. Specifically, we focus on the problem of estimating parameters associated with Lindblad dephasing dynamics, and show that 
entanglement enhancement in the sensitivity (as quantified via quantum Fisher information of the sensor state) arises for estimating the strength of fluctuations in the noise fields' center-of-mass mode. In the presence of strong background fluctuations, such enhancement grows exponentially with the number of probes in the shot-limited regime, when the measurement and reset operations are slow.
This result stands in stark contrast with previously studied scenarios of sensing uncorrelated dephasing noise, where one can prove that entanglement does not lead to an advantage in the signal-to-noise ratio. Our work thus opens a novel pathway towards achieving entanglement-based sensing advantage, which may find applications in characterizing decoherence dynamics of near-term quantum devices. Further, our approach provides a potential quantum-enhanced probe of many-body correlated phases by measuring noise generated by a sensing target. We also discuss realization of our protocol using near-term quantum hardware.
\end{abstract}
\maketitle

{\it Introduction.---}Quantum sensing is one of the most promising applications of near-term quantum systems~\cite{Maccone2011,Cappellaro2017,Treutlein2018,Lloyd2018}, with experimental implementations of, e.g., quantum phase estimation~\cite{Takeuchi2007,Pryde2007,SunLY2019,PanJW2020,Andersen2023}, as well as precision measurements of time~\cite{Schmidt2015,Treutlein2018,YeJun2024}, mechanical motion~\cite{Allcock2019}, electric~\cite{Gleyzes2016,Bollinger2021}, magnetic~\cite{Morton2009,Oberthaler2014}, and gravitational fields~\cite{LIGO2023}. In this context, understanding whether quantum entanglement provides an advantage in estimating an unknown physical quantity is a question of both fundamental importance and tremendous practical interest. Such entanglement-enabled improvement is typically expressed in terms of the asymptotic scaling of the measurement signal-to-noise ratio (SNR) with respect to resource constraints such as the number of atoms used and/or the total interrogation time. For example, for the prototypical example of sensing an unknown parameter $\xi$ coupled to a Hamiltonian $\hat H = \xi \sum _{\ell=1} ^{N} \hat Z_{\ell}$, the optimal strategy is to use the Greenberger–Horne–Zeilinger (GHZ) state, which leads to a $\sqrt{N}$ entanglement advantage in estimating $\xi$. This is known as the Heisenberg limit~\cite{Maccone2004}. 

However, realistic quantum systems are prone to  unwanted environmental noise and decoherence, limiting the performance of quantum sensors in practice. More specifically, no quantum advantage survives in the $N\to \infty$ asymptotic limit when the sensor is subject to uncorrelated Markovian dephasing noise~\cite{Cirac1997,Davidovich2011,Guta_2012,Nori2013,Munro2018,Huelga2018}. In the presence of certain types of spatial or temporal correlations, however, entanglement advantage can survive~\cite{Huelga2014,Duer2017,Sekatski2017,LJiang2018,Duer2020,Kondo2021,Duer2022}. Intriguingly, dissipative dynamics also give rise to a new sensing paradigm of environmental noise estimation~\cite{Monras2008,Nakamura2020,DDobrzanski2023}, which is completely distinct from the standard Hamiltonian case. 
While one can prove that no entanglement advantage is present in the case of sensing independent dephasing noise~\cite{Lupo2017}, whether entanglement helps in sensing more general correlated noise processes remains an open question.

In this Letter, we answer this question positively, by proposing a class of protocols that lead to an entanglement advantage in estimating properties of near maximally-correlated Markovian dephasing noise. We first illustrate the basic mechanism of the new entanglement enhancement using a $2$-qubit example. We then discuss its generalization to multiqubit systems and show that in this setting there can be an \textit{exponential} entanglement advantage in the SNR. Our protocol enables the measurement of the two-point correlation functions of a sensing target, which can be used to design novel quantum-enhanced probes to study interacting many-body quantum systems.
We conclude with a discussion of entanglement enhancement for detecting other forms of Lindbladians and experimental implementation.

{\it Setup: Lindbladian sensing.---}Let us consider a quantum sensor consisting of $N$ qubits that undergoes dynamics described by a Lindbladian $\mathcal{L} _{\xi} $ parametrized by parameter  $ {\xi} $: $({d \hat \rho } / {dt} ) = \mathcal{L} _{\xi} \hat \rho $. While the mechanism discussed here for realizing quantum advantage is general and can be applied to non-Markovian and/or multiaxis noise, 
for concreteness we focus
on sensing Markovian pure-dephasing noise, so that we have
\begin{align}
\label{neq:multi.dephL.def}
\mathcal{L} _{\xi} [C(\xi)]  \hat\rho  = \! 
\frac{\gamma}{2}  \! \sum _{j,\ell =1} ^{N} C _{j\ell}  (\xi) \left( \hat Z_\ell \hat\rho \hat Z_j -\frac{\{ \hat Z_j \hat Z _\ell , \hat \rho\}}{2} \right) 
, 
\end{align}
where $\gamma$ is a constant characterizing the overall decay time scale, 
and $C_{j\ell}  (\xi)$ are matrix elements of a positive semidefinite Hermitian coefficient matrix $C(\xi)$ characterizing the dephasing environment~\cite{reina2002}. Throughout our discussion, we assume $C(\xi) = C(0)+ \xi \Delta C$ depends on the sensing target parameter $\xi$ through a simple linear dependence, without loss of generality~\footnote{If $C(\xi) $ has a more generic functional dependence on $\xi$, then in the context of single-parameter estimation, we can always linearize the function $C (\xi) $ in the vicinity of the estimated value of the parameter $\xi _{0}$ as  $C (\xi) \simeq C (\xi _{0}) + (\xi- \xi _{0}) \partial _{\xi}  C (\xi _{0})$, which would lead to a linear dependence of the dephasing coefficient matrix on $\xi$. }. 
Our goal is to build an unbiased estimator for $\xi$ from measurements on some sensor state following time evolution under $\mathcal{L} _{\xi} $. To illustrate the idea, we assume a typical scenario where the main resource limiting the measurement accuracy is the total protocol time $T$ (see e.g.~\cite{Bollinger2021}), and we are interested in optimizing the amount of extractable information averaged over $T$ 
in the asymptotically long measurement-time (i.e., $T\to \infty$) limit.
The measurement SNR in this case is upper bounded by the quantum Cram{\'{e}}r-Rao bound (CRB)~\cite{Caves1994}: denoting $\Delta _{\xi }$ as the standard deviation of the estimator, we have  
$ (\xi / \Delta _{\xi } ) \le  
\xi \sqrt{ T  \max _{\hat \rho  _{i}, t} \overline{ F _{Q}  } (\hat \rho  _{i}, \mathcal{L} _{\xi};\xi,t) }$, where $\overline{ F _{Q}  } (\hat \rho _{i}, \mathcal{L} _{\xi} ;\xi,t ) $ is defined as the time-averaged quantum Fisher information (QFI) with respect to $\xi$, i.e.,
\begin{align}
\label{neq:chnl.qfi.def}
\overline{ F _{Q}  } (\hat \rho _{i}, \mathcal{L} _{\xi} ;\xi,t )   
\equiv & 8 \lim _{d \xi \to 0}
\frac{ 1- \sqrt{ F (e ^{\mathcal{L} _{\xi} t} \hat \rho  _{i}, e ^{\mathcal{L} _{\xi + d \xi } t} \hat \rho  _{i} ) } }{t d \xi ^{2}} 
. 
\end{align} 
Here, $\rho  _{i}$ is the initial sensor state, and $F (\cdot, \cdot) $ defines the fidelity function between two density operators: 
$F (\hat\rho , \hat\sigma ) =
[ \text{Tr} (\sqrt{\sqrt{\hat\rho } \hat\sigma  \sqrt{\hat\rho }}) ] ^{2} $. 
While prior works have established that a fast-reset strategy achieves the maximal time-averaged QFI over all initial states, $\max  _{ t, \hat \rho  _{i}} \overline{ F _{Q}  } (\hat \rho  _{i}, \mathcal{L} _{\xi} ;\xi,t ) = 
\max _{ \hat \rho  _{i}}\overline{ F _{Q}  } (\hat \rho  _{i}, \mathcal{L} _{\xi} ;\xi,t \! \to\! 0^{+}) $~\cite{Imai2008,Maccone2014,Gefen2025}, proving entanglement advantages requires knowledge of the optimal QFI over protocols that only involve product states, which requires novel techniques because the optimal bounds in~\cite{Imai2008,Maccone2014,Gefen2025} assumed that arbitrary sensor states are available.
In the Supplemental Material~\cite{SI}, we further prove that $\max  _{ t} \overline{ F _{Q}  } (\hat \rho  _{i}, \mathcal{L} _{\xi} ;\xi,t ) = 
\overline{ F _{Q}  } (\hat \rho  _{i}, \mathcal{L} _{\xi} ;\xi,t \! \to\! 0^{+}) $ for pure-dephasing Lindbladians $\mathcal{L} _{\xi}$.

To gain some intuition, we first recall that, in the standard Hamiltonian sensing case, quantum advantage emerges due to an enlarged susceptibility of the sensor GHZ state that is strictly greater than that of any separable sensor state~\cite{Heinzen1994,Oberthaler2014b,Vuletic2022}. This mechanism does not directly extend to the Lindblad sensing case: while a GHZ state does decay faster with the dephasing rate scaling as $N$, such increase in the coherence decay rate also accompanies a degradation in the directly measured signal. In fact, for sensing dephasing rate due to uncorrelated noise, one can show that there is no net gain in SNR when comparing an entangled state with product-state input~\cite{Lupo2017}. For instance, we may consider the problem of estimating a homogeneous single-qubit dephasing rate, $\mathcal{L} _{\xi,\text{1qb}} = \mathcal{L} _{\xi} [C (\xi) =\xi ] $
in Eq.~\eqref{neq:multi.dephL.def}. In this case, the optimal time-averaged QFI for the single-qubit case is obtained by starting in $|+\rangle $: 
\begin{align}
\label{eq:qfi.1qb.deph}
\overline{ F _{Q}  } 
( \left |+\right\rangle 
, \mathcal{L} _{\xi,\text{1qb}} ;\xi,t ) = \gamma / (2\xi) 
.
\end{align} 
The optimal QFI with an $N$-qubit product state is thus given by $N \gamma / (2\xi) $, in accordance with the standard quantum limit (SQL). On the other hand, the optimal sensitivity of measuring $\xi$ with an entangled initial state is saturated by the GHZ state, in which case the maximal time-averaged QFI with respect to $\xi$ can be computed as $
N^{2} \gamma/(2N\xi) $. In this case, the multiqubit GHZ state undergoes a faster dephasing rate $N \xi\gamma$, which has two effects on the time-averaged QFI: the $N^{2}$ factor in the numerator stems from a larger noise susceptibility, while the extra $N$ factor in the denominator is due to reduced signal induced by the enhanced dephasing (cf.~Eq.~\eqref{eq:qfi.1qb.deph}).
Thus the entangled state does not create a net quantum advantage for estimating the original single-qubit dephasing parameter $\xi$. More generally, making use of the no-go theorem in Ref.~\cite{Lupo2017}, one can show that the optimal SNR for estimating any single-qubit Markovian and non-Markovian pure-dephasing rates is given by the SQL.

From the above discussion, we see that the standard mechanism for obtaining a quantum advantage in Hamiltonian estimation does not apply to the case of sensing a single-qubit dephasing rate. This motivates us to explore quantum advantage in pure-dephasing Lindbladians with nontrivial spatial correlation, corresponding to the case where the $C$ matrix in Eq.~\eqref{neq:multi.dephL.def} has nonzero off-diagonal elements. Global noise correlations can occur if the ensemble of sensor qubits is collectively coupled to a dissipative environment, e.g., in trapped-ion systems when the ions experience dephasing due to phase reference error~\cite{Blatt2011} or their collective coupling to the motional modes~\cite{Blatt2015}, in semiconductor and superconducting qubits subject to collective electric~\cite{Yoneda2023} or charge~\cite{McDermott2021} noise, or in an atomic array when the atoms are collectively driven by the same laser beam with nontrivial intensity fluctuations~\cite{Endres2024}. In the context of quantum sensing, as sensor noise measurements can be used to infer the zero- or higher-frequency two-point correlation functions of the environmental fluctuations~\cite{Frank2005,Hirayama2011,Suter2011}, correlated dephasing also naturally arises when sensing fluctuations of a quantum material with nontrivial spatial correlations.

{\it Case study: entanglement advantage in sensing $2$-qubit correlated noise.---}While the quantum advantage we uncover is generally applicable to a variety of correlated dephasing Lindbladians, to illustrate the basic idea, we first consider the case with $2$ qubits subject to a dephasing environment $\mathcal{L} _{\xiCorr,\text{2qb}} = \mathcal{L} _{\xiCorr} [C _{\text{2qb}} (\xiCorr)  ] $ with nearly maximally correlated noise, given by  
$[C _{\text{2qb}} (\xiCorr) ]_{j\ell} = 1 - \xiCorr (1-\delta _{j\ell})
$ with $\xiCorr \ll 1$. Note that $\xiCorr$ arises in off-diagonal elements of $C (\xiCorr)$, signifying that this parameter is related to \textit{spatial correlation} of the dephasing environment. 
The eigenvectors of the Lindbladian $\mathcal{L} _{\xiCorr,\text{2qb}}$ are thus given by outer products 
$\left | \boldsymbol{\alpha} \rangle \langle  \boldsymbol{\beta}  \right |$ spanned by eigenstates of $\hat Z_{1} $ and $\hat Z_{2} $. Here, we use the real vector $\boldsymbol{\alpha} = (\alpha _{1}, \alpha _{2})$ to denote the computational basis states, with 
$\hat Z_{j} \left | \boldsymbol{\alpha} \right \rangle = \alpha _{j} \left | \boldsymbol{\alpha} \right \rangle$. In this case, a generic superposition state of $2$ computational basis vectors 
$| \Psi _{\boldsymbol{\alpha, \beta}} \rangle = (| \boldsymbol{\alpha } \rangle + | \boldsymbol{\beta} \rangle) / \sqrt{2}$ still dephases similarly to the single-qubit case with an effective dephasing rate given by $\gamma _{\boldsymbol{\alpha} ,\boldsymbol{\beta}} =  \frac{\gamma }{4}
(\boldsymbol{\alpha} - \boldsymbol{\beta} ) ^{T} C (\boldsymbol{\alpha} - \boldsymbol{\beta} ) $ (see e.g.~Ref.~\cite{YKLiu2021}), so that the time-averaged QFI with respect to $\xiCorr$ using initial state $| \Psi _{\boldsymbol{\alpha \beta}} \rangle $ is given by 
$\overline{ F _{Q}  } 
( \left |\Psi _{\boldsymbol{\alpha \beta}}\rangle \langle\Psi _{\boldsymbol{\alpha \beta}} \right |, \mathcal{L} _{\xiCorr,\text{2qb}} ;\xiCorr,t ) 
= (\gamma _{\boldsymbol{\alpha} \boldsymbol{\beta}}/2) (\partial \ln \gamma _{\boldsymbol{\alpha} \boldsymbol{\beta}} / \partial \xiCorr) ^{2} $. One can thus show that the optimal initial state that maximizes the QFI per protocol time corresponds to the Bell state 
$| \Psi _{01,10 
} \rangle = (| 01 \rangle + | 10 \rangle) / \sqrt{2}$, which yields $\overline{ F _{Q}  } 
( | \Psi _{01,10}  \rangle  , \mathcal{L} _{\xiCorr,\text{2qb}} ;\xiCorr,t ) 
= \gamma / \xiCorr$. 

We now address the question of whether an entanglement advantage exists in the sensing task of estimating $\xiCorr$ in $\mathcal{L} _{\xiCorr,\text{2qb}}$. For this, we need to compute the optimal time-averaged QFI for all separable states, which can be derived analytically as $\max _{| \phi_{1} \rangle , |\phi_2 \rangle , t} \overline{ F _{Q}  } 
( | \phi_{1} \rangle \! \otimes \! | \phi_{2} \rangle , \mathcal{L} _{\xiCorr,\text{2qb}} ;\xiCorr,t) =\gamma / [{\xiCorr (2- \xiCorr)} ]$~\footnote{Note that the maximal time-averaged QFI over all separable states can always be saturated by a product initial state, by convexity of the quantum Fisher information.} (see~\cite{SI} for details). Noting that $\xiCorr \in [0,1] $ by definition of the Lindbladian in~\eqref{neq:multi.dephL.def}, we have thus shown that using the Bell state 
$| \Psi _{01, 10} \rangle$ generically provides an advantage in estimating $\xiCorr$ parametrizing the 2-qubit Lindbladian for any $\xiCorr<1$. 
Specifically, the entanglement advantage in this case, as quantified by the ratio between optimal sensitivity using entangled versus product states, is 
\begin{align}
\label{eq:qfi.tavg.2qb.ent.adv}
\frac{\max \limits_{\hat \rho  _{i}, t} \overline{ F _{Q}  } (\hat \rho  _{i}, \mathcal{L} _{\xiCorr,\text{2qb}} ;\xiCorr,t) }
{\max \limits_{| \phi_{1}\rangle , | \phi_{2} \rangle , t} \overline{ F _{Q}  } 
( | \phi_{1} \rangle \! \otimes \! | \phi_{2} \rangle , \mathcal{L} _{\xiCorr,\text{2qb}} ;\xiCorr,t) }
= 2 - \xiCorr
. 
\end{align}
In the limit $\xiCorr \ll 1$, the entanglement-advantage ratio takes the maximal value of $2$.

We remark that the entanglement enhancement mechanism in Eq.~\eqref{eq:qfi.tavg.2qb.ent.adv} is completely distinct from the case with standard Hamiltonian sensing. More specifically, in contrast to enhancing the Hamiltonian-estimation QFI using a greater response of entangled states to the target Hamiltonian, in the Lindbladian case we instead make use of the much slower dephasing dynamics of a specific Bell state $| \Psi _{01,10} \rangle $ to enable the entanglement advantage. It is also interesting to note that while it is not physically possible to construct a dephasing Lindbladian where a particular entangled state dephases much faster or slower than all \textit{product} states, as a generic product state will still have nontrivial overlap with the said entangled state, the fact that the QFI monotonically decreases with the entangled state dephasing rate $\gamma _{\boldsymbol{\alpha} \boldsymbol{\beta}}$ means that one can construct an entanglement advantage as long as there exists a unique entangled state that dephases much slower than other eigenstates of the Lindbladian. As we show below, this basic mechanism allows us to construct a class of multiqubit correlated dephasing Lindbladians that hosts a scalable entanglement sensing advantage.

It is worth noting that the entanglement advantage in Eq.~\eqref{eq:qfi.tavg.2qb.ent.adv} persists if the sensing procedure is limited by the total number of measurement runs instead of the probing time. The sensitivity optimized over the number of shots is relevant if the state preparation or detection time is much longer than the time it takes to interrogate the sensing target, which is the case in many current quantum platforms~(see, e.g.,~\cite{Allcock2019}). In this regime, the minimal measurement uncertainty $\Delta {\xiCorr }$ over $M$ shots is set by the maximal quantum Fisher information optimized over the protocol time per shot: $\Delta {\xiCorr } \ge 1/ \sqrt{ M \max _{\hat \rho  _{i}, t} F _{Q}  (\hat \rho  _{i}, \mathcal{L} _{\xiCorr};\xiCorr,t) }$. 
In the $\xiCorr \ll 1$ limit, the maximal QFI per shot over measurement time $t$ is determined by the contribution from $| \Psi _{01,10} \rangle $, and we obtain the same entanglement enhancement as in Eq.~\eqref{eq:qfi.tavg.2qb.ent.adv} (only with the time-averaged QFI $\overline{ F _{Q}  }$ replaced by the QFI per shot,  $F _{Q} $).
Hence, the entanglement advantage applies in both the time-averaged and shot-number-averaged regimes. This feature also distinguishes the entanglement advantage in Eq.~\eqref{eq:qfi.tavg.2qb.ent.adv} from the case of channel estimation discussed in Ref.~\cite{Monras2008}.

{\it Generalization to sensing of correlated multiqubit dephasing.---}While the entanglement advantage discussed in this section does not rely on a specific correlation structure [see discussion following Eq.~\eqref{eq:deph.coeff.nsd}], for concreteness, we focus on target noise correlations with translational invariance. As we discussed, the entanglement enhancement in Eq.~\eqref{eq:qfi.tavg.2qb.ent.adv} crucially relies on the fact that there exists a single entangled state that dephases much slower than any other vectors 
$| \boldsymbol{\alpha } \rangle 
\langle \boldsymbol{ \beta} | $ in the Lindbladian eigenspectrum (excluding diagonal terms that are conserved in the dephasing dynamics). This inspired us to consider the following translationally invariant correlated dephasing Lindbladian (assume total qubit number $N$ is even):
\begin{align} 
\label{eq:lindblad.nqb.def}
\frac{2}{\gamma} \mathcal{L} _{N\mathrm{qb} } & = 
\xiCorr \mathcal{D} [\hat Z_{k=0} ]  + \sum _{k \ne 0} \zeta _{k} \mathcal{D} [\hat Z_k ] 
, 
\end{align} 
where we define $\hat Z_k  =  ({ 1 }/{\sqrt{N}})  \sum  _{j =1} ^{N} e ^{i\frac{2 \pi k}{N}j} \hat Z_j $ ($k=0,1,\ldots,N-1$), and we assume $\zeta _{k}$ are constants satisfying $\zeta _{k} \gg \xi$. 
When $\zeta _{k}=1$, the corresponding coefficient matrix becomes $[C _{N\text{qb}} (\xiCorr)] _{j\ell } =  \delta _{j\ell } - \frac{1-\xiCorr}{N} 
$. In this case, the optimal QFI can be computed as $\overline{ F _{Q}  } 
( | \Psi _{00\ldots , 11\ldots } \rangle  , \mathcal{L} _{\xiCorr,N\text{qb}} ;\xiCorr,t ) 
= N \gamma / 2 \xiCorr $. On the other hand, if we restrict to separable input states $\hat \rho  _{\text{sep}}$, then the optimal time-averaged QFI can be computed as $\max _{\hat \rho  _{\text{sep}}, t} \overline{ F _{Q}  } 
( \hat \rho  _{\text{sep}} , \mathcal{L} _{\xiCorr,N\text{qb}} ;\xiCorr,t) = \gamma / 2 \xiCorr $, giving rise to a factor of $\sqrt{N}$ enhancement in the SNR when using entangled initial states. Here,  entanglement enhancement has the same scaling as the standard Hamiltonian estimation case, although, as we discussed, the underlying mechanism is distinct.

\begin{figure}[t]
    \centering
    \includegraphics[width=\columnwidth]{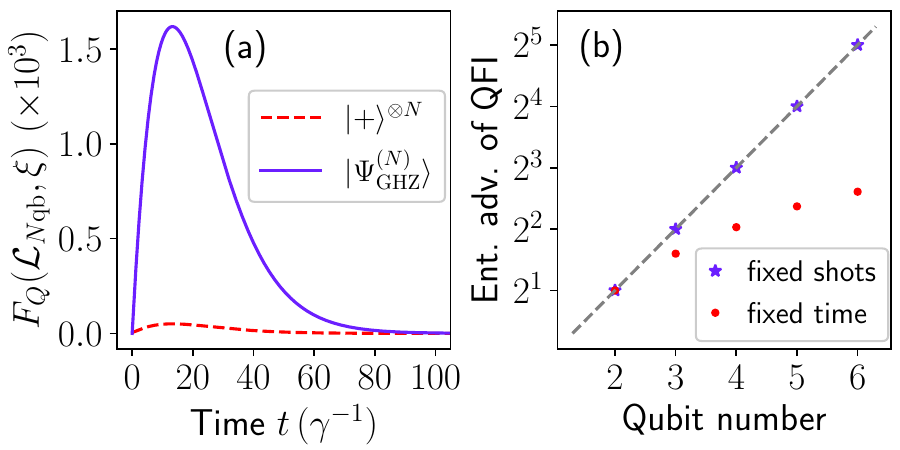}
    \caption{(a) Numerical calculation of the quantum Fisher information (QFI) $F _{Q}  (\hat \rho  _{i}, \mathcal{L} _{N\mathrm{qb} } ;\xi,t)$ for estimating $\xi$ in the multiqubit dephasing [see Lindbladian in Eq.~\eqref{eq:lindblad.nqb.def}] for varying evolution time $t$ per shot, when using the optimal product state $|+\rangle ^{\otimes N}$ versus the optimal entangled state $| \Psi ^{(N)} _{\text{GHZ}} \rangle $ ($N=6$). (b) The entanglement advantage, as quantified by the ratio between the optimal QFI using entangled states and that with only product initial states, plotted against qubit number ($N=2,3,4,5,6$). Both cases of the optimal QFI with fixed total probing time (red dots) and total number of shots (purple stars) are plotted. The dashed line depicts the analytically predicted entanglement advantage for the maximal QFI optimized over number of shots, $2^{N-1}$, which scales 
    exponentially with the qubit number $N$ when fixing the total number of experimental runs, see Eq.~\eqref{eq:qfi.Nqb.ent.adv}.
    }
    \label{fig:ent.adv.lind}
\end{figure}

An even more striking enhancement emerges if the total number of experimental runs, as opposed to the total protocol time, is constrained in the sensing protocol.
In this regime, we care about the optimal QFI per shot, $\max _{\hat \rho  _{i}, t} F _{Q}  (\hat \rho  _{i}, \mathcal{L} _{\xiCorr};\xiCorr,t)$. We further focus on the regime with sufficiently small values of $\xiCorr$, where the lowest nonzero eigenvalue of the Lindbladian in Eq.~\eqref{eq:lindblad.nqb.def} corresponds to coherence decay of the GHZ-type state 
$| \Psi ^{(N)} _{\text{GHZ}} \rangle = (| 0000  \ldots \rangle + | 1111 \ldots  \rangle) / \sqrt{2}$ as 
$\gamma _{_{00 \ldots, 11 \ldots } } = 2 N \xiCorr \gamma $. Similarly to the $2$-qubit case, in the limit $\xiCorr \ll 1/N$, the QFI at times $t \gtrsim 1/\gamma $ is determined by the overlap between the initial state and $| \Psi ^{(N)} _{\text{GHZ}} \rangle $, so that $\lim _{\xiCorr \to 0} \max \limits_{ t} F _{Q}  (\hat \rho  _{i}, \mathcal{L} _{\xiCorr,N\text{qb}} ;\xiCorr,t) \propto   \langle  \Psi ^{(N)} _{\text{GHZ}}  | \hat \rho  _{i} | \Psi ^{(N)} _{\text{GHZ}}    \rangle $~\footnote{As such, we expect any initial states that have $\Omega(1)$ overlap fidelity with $| \Psi ^{(N)} _{\text{GHZ}} \rangle $, including e.g.~highly spin-squeezed states, to exhibit similar form of asymptotic entanglement enhancement in the large qubit number limit ($N\to \infty$).}.  Consequently, the Lindblad sensing exhibits an entanglement advantage that scales exponentially in the number of qubits, as (cf.~\cite{SI})
\begin{align}
& \lim _{\xiCorr \to 0} 
\frac{\max \limits_{\hat \rho  _{i}, t} F _{Q}  (\hat \rho  _{i}, \mathcal{L} _{\xiCorr,N\text{qb}} ;\xiCorr,t) }
{\max \limits_{\hat \rho  _{\text{sep}}, t} F _{Q}  
( \hat \rho  _{\text{sep}} , \mathcal{L} _{\xiCorr,N\text{qb}} ;\xiCorr,t) }
\nonumber\\
\label{eq:qfi.Nqb.ent.adv}
= & \left (  \max _{\hat \rho  _{\text{sep}}}  \left \langle  \Psi ^{(N)} _{\text{GHZ}}  | \hat \rho  _{\text{sep}}| \Psi ^{(N)} _{\text{GHZ}}  \right  \rangle \right ) ^{-1}= 
2^{N-1}
, 
\end{align}
where $\hat \rho  _{\text{sep}}$ denotes an  $N$-qubit separable state. As shown in Fig.~\ref{fig:ent.adv.lind}, numerical calculation of the entanglement advantage for $\xiCorr =0.01$ and up to $N=6$ qubits agrees excellently with the analytical prediction in Eq.~\eqref{eq:qfi.Nqb.ent.adv}. 
The optimal sensitivity set by the QFI is achieved by a POVM $\{ | \Psi ^{(N)} _{\text{GHZ}} \rangle\langle \Psi ^{(N)} _{\text{GHZ}} |, \mathbb{I} - | \Psi ^{(N)} _{\text{GHZ}} \rangle\langle \Psi ^{(N)} _{\text{GHZ}} |\}$, or by measuring each qubit in the Pauli $\hat X$ basis and then extracting the total parity; see End Matter for details.

Given the exponentially large enhancement in the sensitivity, it is natural to ask about the dynamical range of $\xiCorr$ in this sensing task. Here, we define dynamical range of our multiqubit-sensing protocol as the regime where the entanglement advantage in Eq.~\eqref{eq:qfi.Nqb.ent.adv} holds;
from the above discussion, we see that the dynamical range is given by $\xiCorr _{ c } \lesssim N/ 2^{N-1} $. 
We can now view the question of whether $\xiCorr$ lies in the dynamical range as a promise problem: if the uncertainty in estimating $\xi$ over $M$ runs, $\Delta \xiCorr _{N} (M)$, satisfies $\Delta \xiCorr _{N} (M) < \xiCorr _{ c } $, then the promise holds, otherwise it is broken (see the End Matter for details). 
It is interesting to note that, in the case of correlated noise sensing, the promise can be explicitly verified by measuring the coherence of the GHZ state after time $ t \sim 1/ N \xiCorr \gamma$.

We again remark on the drastic distinction between the Lindblad sensing case and the standard Hamiltonian estimation framework: in the standard case of estimating qubit frequency $\xi $ in the Hamiltonian $\hat H = \xi \sum _{\ell=1} ^{N} \hat Z_{\ell}$, one can still assume a promise that the frequency is within the dynamical range $ \xi \ll 1/N $ such that an entanglement advantage holds; however, this promise cannot be explicitly checked, which limits the utility of the GHZ sensing protocol in practice (without adopting an adaptive measurement scheme, see e.g.~\cite{Massar1999,Lukin2014}).

The Heisenberg limit in Hamiltonian sensing is intrinsically a single-shot limit, whereas the dephasing noise sensing problem is intrinsically limited by shot noise, i.e.~in the case of noise sensing, we always have $\Delta \xiCorr \sim T ^{-\frac{1}{2}} $ or $M ^{-\frac{1}{2}}$ , depending on whether we constrain the total protocol time or the number of shots. In the Hamiltonian case, there is a fundamental tradeoff between the achievable sensitivity versus the number of shots implemented; however, in Lindblad noise sensing, increasing the number of shots can lead to a nontrivial enhancement in the protocol sensitivity.

{\it Application: estimating spatial correlations in quantum materials.---}The collective dephasing dynamics in Eq.~\eqref{eq:lindblad.nqb.def} naturally arise when we use an array of qubits to probe noise properties of a spatially extended target~\footnote{For a discussion of how dephasing-noise signals naturally arise in a wide range of qubit-based sensors operating at a finite frequency, see Sec.~I in Supplemental Material~\cite{SI}; cf.~\cite{Preskill2008}.}. The parameter $\xiCorr$ now describes fluctuations in the center-of-mass mode of the noise fields. More specifically, we can generally use a system-environment model to describe the microscopic interaction giving rise to Eq.~\eqref{eq:lindblad.nqb.def}, where the total setup evolves under the Hamiltonian 
$\hat H _{\mathrm{tot} } = 
\frac{1}{2} \sum _{\ell =1} ^{N} 
(\omega_{\ell}
+\hat B _{\ell} )
\hat Z_{\ell}
+ \hat H _{\mathrm{env} } $. In the limit the environment acts as a Gaussian and Markovian bath on the qubits, the dephasing coefficient matrix in Eq.~\eqref{neq:multi.dephL.def} is directly related to correlation functions of the environmental operators as 
\begin{align}
\label{eq:deph.coeff.nsd}
C _{j\ell} = 
\frac{ 1 } {2\gamma } 
\lim \limits _{\omega \to 0}
\overline{\mathcal{S}}_{j\ell} [\omega]
, 
\end{align}
where $\overline{\mathcal{S}}  _{j\ell} [\omega] = 
\frac{1}{2} \int ^{\infty}_{ - \infty  } 
\langle  \{  \hat B  _{j}  (s) ,
\hat B  _{\ell} (0) 
\}  \rangle e^{i \omega s } ds  $ are the symmetrized noise spectral densities of $\hat B  _{j}$, with $\hat B  _{j}  (s) \equiv e ^{i \hat H _{\mathrm{env} } s} 
\hat B  _{j}e ^{-i \hat H _{\mathrm{env} } s}  $ denoting the interaction-picture environmental noise operators. Taking the spatial Fourier transform of the noise operators, $\hat B  _{k} = ({ 1 }/{\sqrt{N}})  \sum  _{j =1} ^{N} e ^{i\frac{2 \pi k}{N}j} \hat B  _{j} $, we see that the correlated noise environment in Eq.~\eqref{eq:lindblad.nqb.def} corresponds to the case where $\overline{\mathcal{S}} _{k=0} [\omega] / \overline{\mathcal{S}} _{k\ne 0} [\omega] = \xiCorr$. As shown in Supplemental Material~\cite{SI}, the exponential entanglement enhancement and the associated analysis of dynamical range are applicable as long as the noise spectral density in momentum space is gapped, with $\overline{\mathcal{S}} _{k=0} [\omega] / (\min_{k'\ne 0} \overline{\mathcal{S}} _{k'} [\omega]) = \xiCorr$. 
In contrast, in the time-limited regime, the entanglement advantage in estimating the collective $\hat Z_{k=0} $ noise in Eq.~\eqref{eq:lindblad.nqb.def} holds regardless of the values of other nuisance $\hat Z_{k\ne 0} $ terms. We note that Eq.~\eqref{eq:deph.coeff.nsd} also applies if the environment consists of classical stochastic quantities, in which case $\mathcal{S} _{j\ell} [\omega]$ correspond to the noise spectral densities of those classical fluctuations. 

While our protocol can extract information about zero-frequency noise spectral densities of the environment (i.e., the sensing target), our approach is not limited to measuring zero-frequency or low-frequency noise.
By applying standard dynamical decoupling or spin-locking sequences, we can also extend Eq.~\eqref{eq:deph.coeff.nsd} to measuring spectral properties of the environmental noise at nonzero frequencies~\cite{Hirayama2011,Suter2011}. Intriguingly, Eq.~\eqref{eq:lindblad.nqb.def} naturally arises if the environment is fully scrambled except for the center-of-mass (i.e., $k=0$) mode; see the End Matter for details on the practical relevance of Eq.~\eqref{eq:lindblad.nqb.def}.

{\it Discussion and outlook.---}In this Letter, we uncover a new mechanism for obtaining entanglement advantage in sensing spatially-correlated Markovian noise. When the total protocol time is fixed, we find a Heisenberg-type scaling in the time-averaged QFI for estimating a small deviation of sensor dynamics from maximally correlated dephasing; 
such an entanglement advantage even scales exponentially when the total number of experimental runs is constrained. While super-Heisenberg scaling has been explored in the context of Hamiltonian sensing as well, almost all known examples of larger-than-Heisenberg entanglement advantage involve a sensor Hamiltonian that can generate entanglement during time evolution~\cite{Geremia2007,Braunstein2008,Maccone2011}. In contrast, the correlated dephasing dynamics we consider in this work cannot create any entanglement.
In the End Matter and Supplemental Material~\cite{SI}, we also discuss extensions of the entanglement advantages in Eq.~\eqref{eq:qfi.Nqb.ent.adv} to other Lindbladians and to non-Markovian targets.

While entanglement with ancillae is not required to achieve the optimal QFI for estimating the correlated dephasing Lindbladians considered in our work, ancillae can help estimating multiaxis noise or amplitude decay in the presence of nuisance terms~\cite{Sekatski2022,Gefen2025}. 
A natural question to ask is how generic decoherence processes would impact the quantum sensing advantage in our work; we expect the exponential advantage in Eq.~\eqref{eq:qfi.Nqb.ent.adv} to survive, as long as those spurious terms preserve the condition $\overline{\mathcal{S}} _{k=0} [\omega] / (\min_{k'\ne 0} \overline{\mathcal{S}} _{k'} [\omega]) = \xiCorr$. We leave a detailed analysis of those questions to future work. 
Another related type of sensing tasks involves channel estimation~\cite{DDobrzanski2023}. 
An intriguing direction for future investigation is whether an exponential advantage emerges in more general channel estimation problems.

As mentioned, our framework [cf.~Eq.~\eqref{eq:deph.coeff.nsd}] can be directly generalized to study non-Markovian noise of the environment, with the help of standard qubit noise spectroscopy techniques (see, e.g.,~\cite{Oliver2011,Oliver2012,Viola2017,Oliver2020,deLeon2022,deLeon2024,DuJiangfeng2024,Degen2024,deLeon2024multi,Kolkowitz2024}). An outstanding challenge is whether one can obtain a quantum advantage in probing a non-Markovian environment. While we do not expect any entanglement-based enhancement in probing spatially uncorrelated dephasing noise, finding the optimal strategy to reconstruct environmental noise properties with both nontrivial spatial and temporal correlations remains an open question.
It is also interesting to explore possible extensions of our results to multi-parameter estimation. Previous works have shown that entanglement can enable an exponential advantage in the number of samples required to learn Pauli channel parameters (see e.g.~Refs.~\cite{Jiang2024,Minev2024}). It is thus intriguing to ask whether our results can be used to enhance learning algorithms for general non-Pauli dephasing channels.
We leave this to future work.

{\it Acknowledgments.---}We thank Lorenza Viola, Gideon Lee, Sisi Zhou, Rafa{\l{}} Demkowicz-Dobrza{\'{n}}ski, and Liang Jiang
for helpful discussions. 
We thank Haowei Shi and Quntao Zhuang for pointing out an error in the proof of the optimal time-averaged QFI in an earlier version of the Supplemental Material. 
Y.-X.W.~acknowledges support from a QuICS Hartree Postdoctoral Fellowship. 
J.B.~acknowledges support from the Harvard Quantum Initiative. 
J.B.~and A.V.G.~were supported in part by AFOSR MURI, ONR MURI, DARPA SAVaNT ADVENT, NSF STAQ program, NSF QLCI (award No.~OMA-2120757), DoE ASCR Quantum Testbed Pathfinder program (awards No.~DE-SC0019040 and No.~DE-SC0024220), ARL (W911NF-24-2-0107), and NQVL:QSTD:Design:FTL.  A.J.B.~acknowledges support from the NRC Research Associateship Program at NIST. 
C.O.~was supported by the National Research Foundation of Korea Grants (No.~RS-2024-00431768 and No.~RS-2025-00515456) funded by the Korean government (Ministry of Science and ICT (MSIT)) and the Institute of Information \& Communications Technology Planning \& Evaluation (IITP) Grants funded by the Korean government (MSIT) (No.~RS-2024-00437284, No.~IITP-2025-RS2025-02283189 and No.~IITP-2025-RS-2025-02263264) and by Global Partnership Program of Leading Universities in Quantum Science and Technology (RS-2025-08542968) through the National Research Foundation of Korea (NRF) funded by the Korean government (Ministry of Science and ICT(MSIT)).
J.B.~and A.V.G. also acknowledge support from the U.S.~Department of Energy, Office of Science, National Quantum Information Science Research Centers, Quantum Systems Accelerator (QSA) and from the U.S.~Department of Energy, Office of Science, Accelerated Research in Quantum Computing, Fundamental Algorithmic Research toward Quantum Utility (FAR-Qu).

\onecolumngrid

\vspace{.3in}
\begin{center}
\textbf{End Matter}
\end{center}
\vspace{.1in}

\twocolumngrid

In this End Matter, we provide more detail regarding optimal measurements achieving the QFI in detecting the correlated noise parameter $\xi$ in Eq.~\eqref{eq:lindblad.nqb.def}, dynamical range of the exponential entanglement advantage [cf., Eq.~\eqref{eq:qfi.Nqb.ent.adv}], practical relevance of correlated noise structures enabling such exponential advantage and their non-Markovian extensions, and generalization of the entanglement advantages discussed in the main text to other Lindbladians. \\

{\it Optimal measurements achieving the QFI for correlated noise metrology.---}In the main text, we discuss an exponential entanglement advantage in the QFI for estimating $\xi$ in the shot-limited regime [see Eq.~\eqref{eq:qfi.Nqb.ent.adv}]. 
The optimal sensitivity set by the QFI is achieved by a POVM $\{ | \Psi ^{(N)} _{\text{GHZ}} \rangle\langle \Psi ^{(N)} _{\text{GHZ}} |, \mathbb{I} - | \Psi ^{(N)} _{\text{GHZ}} \rangle\langle \Psi ^{(N)} _{\text{GHZ}} |\}$, or by measuring each qubit in the Pauli $\hat X$ basis and then computing the measurement outcome correlations to extract the total parity. 
More specifically, the entangled-basis measurements can be implemented by, e.g., first performing a time reversal of the unitary that transforms the computational basis state $\ket{0}^{\otimes n}$ into the initial sensor GHZ state $| \Psi ^{(N)} _{\text{GHZ}} \rangle $, followed by single-qubit computational basis measurements. This strategy, also known as a many-body Loschmidt-echo measurement~\cite{Pezze2016}, is robust to detection noise and does not introduce any control overhead beyond what is already required in the entanglement strategy. As the optimal measurement scheme only involves two outcomes, the finite-sample performance of this estimator is the same as a two-variable Bernoulli process.\\

{\it Detail on dynamical range of the exponential entanglement advantage in Eq.~\eqref{eq:qfi.Nqb.ent.adv}.---}In the main text, we define the dynamical range $\xiCorr _{ c }$ of the correlated noise sensing problem as the regime where the entanglement advantage in Eq.~\eqref{eq:qfi.Nqb.ent.adv} holds.
Within this range, the uncertainty of measurement over $M$ runs can be computed from the QFI as 
\begin{align}
\Delta \xiCorr _{N} (M) = \left [ M \max \limits_{ t} F _{Q}  (| \Psi ^{(N)} _{\text{GHZ}} \rangle , \mathcal{L} _{\xiCorr,N\text{qb}};\xiCorr,t)\right ] ^{-\frac{1}{2}}
,
\end{align} 
which can be straightforwardly shown to scale as $\Delta \xiCorr _{N} (M) \propto \xiCorr/ \sqrt{ M} $. 
We stress that the dynamical range is a completely distinct physical quantity from the entanglement advantage, and generally, their values do not need to be related. If the measurement uncertainty now satisfies 
$\Delta \xiCorr _{N} (M) < \xiCorr _{ c } $, then the measurement contributes towards gaining more information about $\xiCorr $; conversely, if the measurement uncertainty is greater than the promised value $\Delta \xiCorr _{N} (M) > \xiCorr _{ c } $, then our measurement does not improve the measurement uncertainty beyond  the original promise. 
We also note that if the target parameter $\xiCorr$ lies beyond the dynamical range ($\xiCorr > \xiCorr _{ c }$), a nontrivial entanglement advantage in estimating $\xiCorr$ still survives, but it will scale subexponentially with the sensor size $N$.
\\

{\it Practical relevance of Eq.~\eqref{eq:lindblad.nqb.def} and extension of the exponential entanglement advantage to non-Markovian targets.---}As discussed in the main text, the noise correlation structure in Eq.~\eqref{eq:lindblad.nqb.def} naturally arises if the environment is fully scrambled in almost all the degrees of freedom, except the center-of-mass (i.e., $k=0$) mode. The latter may happen if the sensing target is a many-body system subject to conservation law of fluctuations in the center-of-mass mode. Alternatively, Eq.~\eqref{eq:lindblad.nqb.def} naturally describes dephasing noise environments supporting the decoherence-free subspace $\{| 0000  \ldots \rangle , | 1111 \ldots  \rangle\}$,  which may occur in trapped ions due to the distinct timescales of the different $\hat B  _{k}$ noise modes~\cite{Blatt2011} or in specially engineered superconducting qubits~\cite{Siddiqi2016}.
More generally, in the shot-limited regime, if the spatiotemporal bath correlators satisfy $\braket{\hat B  _{j}  (s)  \hat B  _{\ell} (0)  } = C _{j\ell} (\xi) \cdot S_{0}(s) $, then a mechanism similar to Eq.~\eqref{eq:qfi.Nqb.ent.adv} persists. In this case, the temporal correlation $S_{0} (\cdot)$ can modify the optimal sensing time, but any nontrivial temporal correlation would not remove the exponential advantage as long as $C _{j\ell} (\xi)$ hosts the many-body gapped spatial correlation pattern. Such correlated non-Markovian environments can include a variety of naturally occurring noise sources such as electrical noise and nuclear spin baths in semiconductor quantum-dot qubits~\cite{Yoneda2023} and photon shot noise in superconducting qubits~\cite{Oliver2020}, provided the spatial noise correlation supports the many-body gap structure. Further, in the shot-limited regime, the entangled sensing protocol discussed after Eq.~\eqref{eq:qfi.Nqb.ent.adv} can also be directly applied to the non-Markovian case.\\

{\it Entanglement advantages in estimating other correlated Lindbladians.---}The entanglement enhancement discussed in the main text extends to estimating correlations in other Lindbladians beyond qubit dephasing, including e.g.~bosonic loss, gain, and random displacements (see Supplemental Material~\cite{SI} for details). In those latter cases, the entanglement advantage scales at most linearly in the number of modes in both time-limited and shot-limited regimes due to the Gaussian nature of the target dynamics. In contrast, the mechanism~\eqref{eq:qfi.Nqb.ent.adv} for enabling exponential entanglement advantage directly generalizes to bosonic dephasing (under an appropriate energy constraint)
\begin{align}
\label{neq:multi.dephL.boson}
\mathcal{L} _{\text{boson},\xi} [C(\xi)]  \hat\rho  = \! 
\frac{\gamma}{2}  \! \sum _{j,\ell =1} ^{N} C _{j\ell}  (\xi) \left( \hat n_\ell \hat\rho \hat n_j -\frac{\{ \hat n_j \hat n _\ell , \hat \rho\}}{2} \right) 
, 
\end{align}
where we denote by $\hat n_{j}$ the number operator of the $j$th bosonic mode, as long as the dephasing noise correlation matrix $C _{j\ell}  (\xi)$ supports a many-body gap in the Fock basis.
It will be interesting to consider generalizations of the exponential entanglement advantage beyond pure-dephasing Lindbladians, which we leave to future work. \\

\clearpage

\onecolumngrid
\begin{center}
\textbf{\large Supplemental Material: Exponential entanglement advantage in sensing correlated noise}
\end{center}
\vspace{.1in}
\twocolumngrid
\setcounter{section}{0}
\setcounter{equation}{0}
\setcounter{figure}{0}
\setcounter{page}{1}
\renewcommand{\theequation}{S\arabic{equation}}
\renewcommand{\thefigure}{S\arabic{figure}}

In this Supplemental Material, we provide details on justification of pure-dephasing noise in typical qubit-based sensors, proof for the optimality of a fast-reset strategy in the time-limited regime, generalizations and robustness of exponential entanglement advantage in sensing noise in the center-of-mass mode in the presence of a many-body Lindbladian dephasing-rate gap, and entanglement advantage in estimating various bosonic Lindbladians.

\tableofcontents 

\appendix

\begingroup
\renewcommand{\thesubsection}{I}
\subsection{Justification of pure-dephasing noise in finite-frequency qubit sensors}
\endgroup 

While the entanglement advantage discussed in the main text can be extended to more general forms of Lindbladians beyond the pure-dephasing form, for practical qubit-based noise sensors operating at a finite (e.g., microwave) frequency, the pure-dephasing Lindbladian usually provides an excellent approximation of the dominant noise which the sensor qubit is sensitive to. Here, we provide a physical justification for the pure-dephasing approximation that is valid for a wide range of state-of-the-art qubit noise sensors, including solid-state defect spin qubits (such as nitrogen vacancy centers in diamond~\cite{deLeon2022,deLeon2024}), semiconductor quantum-dot spin qubits described by central-spin models~\cite{ZhangWenxian2019,ZhangWenxian2026}, and superconducting flux-tunable qubits~\cite{Oliver2011,Oliver2012,Oliver2020}. Specifically, the dephasing of a finite-frequency sensor qubit is due to low-frequency noise, whereas its population loss and gain is due to environmental fluctuations that are near resonance with the qubit frequency (see, e.g.,~\cite{Schon2005}). For a sensor qubit operating at microwave or higher frequencies, this means that dephasing typically occurs at a much faster timescale than qubit population loss and heating, naturally introducing a biased noise structure for the sensor qubits~\cite{Preskill2008}.

We start from a generic microscopic model describing a set of sensor qubits coupled to its environment:
\begin{align}
\hat H _{\mathrm{tot} } = 
\frac{1}{2} \sum _{\ell =1} ^{N} \omega_{\ell} \hat Z_{\ell} 
+ \hat H _{\mathrm{env} } + \hat H _{\mathrm{int} } 
,
\end{align}
where we allow both longitudinal (i.e., qubit-population preserving) and transverse (i.e., qubit-population non-preserving) noise couplings between the sensor qubit and its environment, as 
\begin{align}
\hat H _{\mathrm{int} } = 
\frac{1}{2} \sum _{\ell =1} ^{N} 
(\hat B _{z,\ell} \hat Z_{\ell} 
+ \hat B _{x,\ell}  \hat X _{\ell}
+ \hat B _{y,\ell}  \hat Y _{\ell})
. 
\end{align}
In the interaction picture with respect to the intrinsic qubit and environmental Hamiltonians $\hat H _{0} = \sum _{\ell =1} ^{N} (\omega_{\ell}/2) \hat Z_{\ell}  + \hat H _{\mathrm{env} }$, the sensor-environment interaction Hamiltonian reads
\begin{align}
&  \hat H _{\mathrm{int} } (t) = \frac{1}{2} \sum _{\ell =1} ^{N}  
[\hat B _{z,\ell}  (t)
\hat Z_{\ell} 
\nonumber \\
& \quad + \hat B _{-,\ell} (t) \hat \sigma  _{+,\ell} e ^{i\omega_{\ell} t }
+ \hat B _{+,\ell} (t)  \hat \sigma  _{- ,\ell} e ^{- i\omega_{\ell} t } ]
,
\end{align}
 with $\hat B  _{j}  (s) \equiv e ^{i \hat H _{\mathrm{env} } s} 
\hat B  _{\alpha, j}e ^{-i \hat H _{\mathrm{env} } s}  $ ($\alpha = x,y,z$) denoting the interaction-picture environmental noise operators, $\hat B _{\pm,\ell} (t) = \hat B _{x,\ell}  \pm i \hat B _{y,\ell} $ and $ \hat \sigma  _{\pm,\ell} = (\hat X _{\ell}\pm i \hat Y _{\ell})/2$ denoting the effective raising and lowering noise and qubit operators. For convenience, we introduce the quantum noise spectral densities as~\cite{QN.RMP2010} 
\begin{align}
\label{seq:nsd.gen.def}
\mathcal{S } ^{(\alpha, \beta)}_{ j\ell} [\omega ] = 
\int ^{\infty}_{ - \infty  } 
\langle \hat B  _{\alpha,j}  (s)  
\hat B  _{\beta, \ell} (0)  \rangle 
e^{i \omega s } ds  
.
\end{align} 
The symmetrized noise spectral densities introduced in the main text are thus given by $\overline{\mathcal{S }} _{ j\ell} [\omega ]  = (\mathcal{S } _{ j\ell} [\omega ] + \mathcal{S } _{\ell j} [- \omega ])/2$.

Under the standard Born-Markov approximation~\cite{oqs2002book}, we can integrate out the environmental dynamics and obtain a quantum master equation for the sensor dynamics (assuming $ |\omega _{j} - \omega _{\ell} | \ll \omega _{j} \simeq \omega _{\ell}$):
\begin{align}
\frac{d}{dt} \hat \rho = & \sum _{j,\ell =1} ^{N} 
\gamma ^{(z)}_{j\ell}  \left( \hat Z_\ell \hat\rho \hat Z_j -\frac{1}{2} \{ \hat Z_j \hat Z _\ell , \hat \rho\}\right) 
\nonumber \\
&+ \gamma ^{(-)}_{j\ell}  \left( \hat \sigma _{- ,\ell} \hat\rho  \hat \sigma _{+ ,j} -\frac{1}{2} \{  \hat \sigma _{+ ,j} \hat \sigma _{- ,\ell}  , \hat \rho\}\right) 
\nonumber \\
&+ \gamma ^{(+)}_{j\ell}  \left( \hat \sigma _{+ ,\ell} \hat\rho  \hat \sigma _{- ,j} -\frac{1}{2} \{  \hat \sigma _{- ,j} \hat \sigma _{+ ,\ell}  , \hat \rho\}\right) 
, 
\end{align}
where the coefficients $\gamma ^{(z)}_{j\ell} $ and $\gamma ^{(\pm)}_{j\ell} $ are given by the corresponding environmental noise spectral densities
\begin{align}
\gamma ^{(z)}_{j\ell} & = \frac{1}{4}\lim \limits _{\omega \to 0} \overline{\mathcal{S}} ^{(zz)}_{j\ell} [\omega]  
, \\
\gamma ^{(-)}_{j\ell} & = \mathcal{S } ^{(-+)}_{ j\ell} [+\omega _{j} ] 
, \\
\gamma ^{(+)}_{j\ell} & = \mathcal{S } ^{(+-)}_{ j\ell} [- \omega _{\ell} ] 
.
\end{align}
Thus, under the Born-Markovian approximation, the leading-order dephasing effect originates from low-frequency noise. In contrast, the population non-conserving decoherence effects are coupled to fluctuations near the qubit frequency, as these processes must deposit or absorb energy in units of qubit frequency into/from the environment. 

From the perspective of sensor engineering, it is straightforward to design qubits that are much more sensitive to $T_{2}$-type noise by engineering a large energy or potential barrier for the qubit (also known as biased-noise qubits~\cite{Preskill2008,Terhal2009}), whereas it is much more challenging to realize qubits that are only sensitive to $T_{1}$ processes. 
We also note that our sensing protocol does not require any static coupling between the probe qubits and can operate with a weak local sensor-qubit Rabi rate, as long as the proper entangled sensor state can be prepared before the onset of signal accumulation.
\\

\begingroup
\renewcommand{\thesubsection}{II}
\subsection{Proof of optimal time-averaged quantum Fisher information (QFI) for estimating pure-dephasing Lindbladians}
\endgroup

In this section, we prove that a fast-reset strategy with appropriate initial sensor states always achieves the optimal time-averaged quantum Fisher information (QFI) in estimating the pure-dephasing Lindbladians considered in the main text. We then apply the general result to the multiqubit Lindbladians $\mathcal{L} _{\xi} [C _{N\text{qb}} (\xiCorr)]$ considered in the main text and explicitly derive the optimal time-averaged QFI. \\

\subsubsection{General formula and statement of result}

We first recall that the pure-dephasing Lindbladian $ \mathcal{L} _{\xi} [C (\xiCorr)] $ can be written in the following form (see Eq.~\eqref{neq:multi.dephL.def} in the main text):
\begin{align}
\label{seq:multi.dephL.def}
\mathcal{L} _{\xi} [C(\xi)]  \hat\rho  = \! 
\frac{\gamma}{2}  \! \sum _{j,\ell =1} ^{N} C _{j\ell}  (\xi) \left( \hat Z_\ell \hat\rho \hat Z_j -\frac{\{ \hat Z_j \hat Z _\ell , \hat \rho\}}{2} \right) 
, 
\end{align}
where $\gamma$ denotes a characteristic dephasing rate, 
and $C_{j\ell} (\xi)$ are matrix elements of the positive semidefinite Hermitian coefficient matrix $C(\xi)$ characterizing the environmental noise correlations. As discussed in the main text, we assume $C(\xi) = C(0)+ \xi \Delta C$, which can be viewed as a first-order expansion in $\xi$ when the dependence $C(\xi)$ is more general than a linear function. For notational simplicity, we also define the superoperator corresponding to the differential Lindbladian with respect to $\xiCorr$ as
\begin{align}
\label{seq:diffL.def}
\partial _{\xiCorr} \mathcal{L} = \frac{\gamma}{2}  \! \sum _{j,\ell =1} ^{N} (\Delta C) _{j\ell}  \left( \hat Z_\ell \hat\rho \hat Z_j -\frac{\{ \hat Z_j \hat Z _\ell , \hat \rho\}}{2} \right) 
.
\end{align}
For any initial state $\hat \rho _{i}$, we can shift the jump operators by arbitrary constants $\hat Z _{\ell} \to \hat Z' _{\ell} = \hat Z _{\ell}  - \text{Tr} (\hat Z _{\ell} \hat \rho _{i}) $ without changing the total Lindbladian $\mathcal{L} _{\xi} [C(\xi)] $. Thus, without loss of generality, we assume that we have redefined the jump operators such that their expectation values in the initial state are zero in the rest of this section.

In Eq.~\eqref{neq:chnl.qfi.def} in the main text, we introduce the time-averaged QFI $\overline{ F _{Q}  } (\hat \rho  _{i}, \mathcal{L} _{\xi} ;\xi,t )$, which quantifies the maximal information extractable about $\xi$ by measuring the time evolution of $\hat \rho  _{i}$ under $\mathcal{L} _{\xi}$ for time $t$. This quantity can be formally defined using the fidelity function, or equivalently the Bures metric $d_{\mathrm{Bures}}$, as 
\begin{align}
& \overline{ F _{Q}  } (\hat \rho  _{i}, \mathcal{L} _{\xi} ;\xi,t )   
\equiv 8 \frac{ 1- \sqrt{F (e ^{\mathcal{L} _{\xi} t} \hat \rho  _{i}, e ^{\mathcal{L} _{\xi + d \xi } t} \hat \rho  _{i} ) } }{t d \xi ^{2}} 
\nonumber \\
\label{seq:lindblad.qfi.def}
= &4  \frac{ d ^{2}_{\mathrm{Bures}} (e ^{\mathcal{L} _{\xi} t} \hat \rho  _{i}, e ^{\mathcal{L} _{\xi + d \xi } t} \hat \rho  _{i} ) }{t d \xi ^{2}} 
. 
\end{align} 
When the total protocol time is fixed, the optimal sensitivity is given by the time-averaged QFI for any given initial state $\hat \rho  _{i}$ maximized over all evolution times $t$ as (cf.~main text)
\begin{align}
\label{seq:qCRB.time.avg}
\left .\Delta  {\xi } \right |_{\hat \rho  _{i}} \ge 1/ \sqrt{ T  \max _{ t} \overline{ F _{Q}  } (\hat \rho  _{i}, \mathcal{L} _{\xi};\xi,t) }
,
\end{align}
where $\left .\Delta  {\xi } \right |_{\hat \rho  _{i}} $ denotes the standard error of the estimated $\xi $ in the long-measurement-time limit. By convexity of the QFI, we only need to consider pure initial states in the following discussion.

Our goal is to prove the following equivalence between the optimal and the fast-reset time-averaged QFI:
\begin{align}
\max _{ t} \overline{ F _{Q}  } (\ket{\phi _{i}}, \mathcal{L} _{\xi};\xi,t) 
= \lim _{t\to 0^{+}} \overline{ F _{Q}  } (\ket{\phi _{i}}, \mathcal{L} _{\xi};\xi,t) 
.
\end{align}

\subsubsection{Fast-reset limit of the time-averaged QFI}

It is straightforward to see that the QFI in the regime $\gamma t \to 0 ^{+} $ in Eq.~\eqref{seq:lindblad.qfi.def} can be evaluated analytically. For convenience, we rewrite the Lindbladian coefficient matrix $C(\xi)$ in terms of a factor matrix $g (\xi)$ with row vectors $g_{\ell}   (\xi) $, $[g (\xi)] ^{T} = (g_{1}^{T}  (\xi), \ldots , g_{N}^{T}  (\xi) ) $:
\begin{align}
C(\xi)  = g ^{\dag} (\xi) g (\xi) 
= \sum _{\ell=1} ^{N} g _{\ell}^{\dag} (\xi) g _{\ell} (\xi) 
.
\end{align}
The factor matrix $g (\xi)$ is uniquely defined up to a gauge degree of freedom, as $g' (\xi) = W(\xi) g (\xi)$ for an arbitrary unitary matrix $W(\xi)$ will also satisfy the above condition. Given any specific initial state $\hat \rho _{i} = \left | \phi _{i} \rangle \langle  \phi _{i} \right |$, and using the gauge degree of freedom in defining $g (\xi) $, we can always choose a gauge such that the following orthogonality condition is satisfied: 
\begin{align}
\label{seq:diff.psi.ortho}
\sum _{m,n=1} ^{N} g _{\ell m} (\xi) g ^{*} _{j n} (\xi) \langle \phi _{i} | \hat Z ' _{n} \hat Z ' _{m} | \phi _{i}   \rangle  \propto \delta _{\ell  j }
. 
\end{align}
Under this gauge, we can write the sensor state after an infinitesimal time in terms of the corresponding orthogonal basis states $\{ \ket{\psi ^{\perp}  _{\ell} } = \frac{1}{\sqrt{p _{\ell}}}\sum _{m} g _{\ell m} (\xi)  \hat Z ' _{m} \ket{\phi _{i} }\} $ and additional contributions generated by non-Hermitian-like propagation: 
\begin{align}
& \hat \rho _{\xi} (dt ) = \hat \rho _{i} + \frac{\gamma dt}{2}  \! \sum _{j,\ell =1} ^{N} \left ( g ^{\dag} (\xi) g (\xi)  \right ) _{j\ell} 
\nonumber \\
& \quad \times \left( \hat Z' _\ell \left | \phi _{i} \rangle \langle  \phi _{i} \right | \hat Z ' _j -\frac{1}{2} \{ \hat Z' _j \hat Z'  _\ell , \hat \rho _{i} \} \right) + O(dt^{2})
\nonumber \\
= & \hat \rho _{i} +  \frac{\gamma dt}{2}  \! \sum _{\ell =1} ^{N} p _{\ell} \left | \psi ^{\perp}  _{\ell} \rangle \langle \psi ^{\perp}  _{\ell}  \right | 
\nonumber \\
& -  \frac{\gamma dt}{4} \sum _{j,\ell =1} ^{N} C _{j\ell}  (\xi)  \{  \hat Z' _j \hat Z'  _\ell , \hat \rho _{i} \}
\nonumber \\
& + O(dt^{2})
.
\end{align}
From $\partial _{\xi} C(\xi) =  \Delta C(\xi)$, we have 
\begin{align}
\label{seq:DeltaC.tog}
& \Delta C (\xi) = [\partial _{\xi} g ^{\dag} (\xi) ] g (\xi) + g ^{\dag}  (\xi) \partial _{\xi} g (\xi)   
,
\end{align}
so that we can write the partial differential density matrix with respect to $\xi $ in the $t \to 0 ^{+}$ limit as
\begin{align}
& \partial _{\xi} \hat \rho _{\xi} (dt ) = \frac{\gamma dt}{2}  \! \sum _{j,\ell =1} ^{N} \left ( [\partial _{\xi} g ^{\dag} (\xi) ] g (\xi) + g ^{\dag}  (\xi) \partial _{\xi} g (\xi)   \right ) _{j\ell} 
\nonumber \\
& \times \left( \hat Z' _\ell \left | \phi _{i} \rangle \langle  \phi _{i} \right | \hat Z ' _j -\frac{1}{2} \{ \hat Z' _j \hat Z'  _\ell , \hat \rho _{i} \} \right) + O(dt^{2})
.
\end{align}
From Eq.~\eqref{seq:diff.psi.ortho}, we see that
\begin{align}
\langle \psi ^{\perp}  _{\ell}  | \partial _{\xi} \hat \rho _{\xi} (dt ) | \psi ^{\perp}  _{j}  \rangle \propto \delta _{\ell j }
.
\end{align}

We can now explicitly compute the time-averaged QFI in the $t\to 0 ^{+}$ limit with respect to $\xi$ in terms of the symmetric logarithmic derivative (SLD; see, e.g.,~\cite{Caves1994}). Given any density operator with eigenvalue decomposition $\hat \rho = \sum _{\ell} p _{\ell}  \left | r _{\ell} \rangle \langle  r _{\ell} \right | $, we define a \textit{symmetric inverse} superoperator with respect to $\hat \rho$ of an operator $\hat O$ as 
\begin{align}
\mathcal{R} ^{-1} _{\hat \rho} \hat O \equiv 2 \sum _{j,\ell; } [|\langle r _{j} | \hat O | r _{\ell} \rangle | ^{2}/(p _{j}+ p _{\ell} )] | r _{j} \rangle \langle r _{\ell} | 
.
\end{align}
The SLD of a state $\hat \rho $ with respect to $\xi$ is thus defined as $\mathcal{R} ^{-1} _{\hat \rho}(\partial _{\xi} \hat \rho  )$.
The eigenbasis of the SLD also provides the measurement basis that saturates the optimal sensitivity as given in Eq.~\eqref{seq:qCRB.time.avg}. We thus have
\begin{align}
& \lim _{t\to 0 ^{+}} \left (
\overline{ F _{Q}  } (\hat \rho  _{i}, \mathcal{L} _{\xi};\xi,t)  \right )
\nonumber \\
= & \lim _{t\to 0 ^{+}} \left [\frac{\text{tr} (\partial _{\xiCorr} \mathcal{L} _{\xiCorr} ( e ^{\mathcal{L} _{\xi} t} \hat \rho  _{i} ) 
\mathcal{R} ^{-1} _{e ^{\mathcal{L} _{\xi} t} \hat \rho  _{i} } (\partial _{\xiCorr} \mathcal{L} _{\xiCorr} ( e ^{\mathcal{L} _{\xi} t} \hat \rho  _{i} ) ))}{t}  \right ]
\nonumber \\
= & 2 \sum _{j, \ell =1}^{N}  \frac{|\langle \psi ^{\perp}  _{\ell}  | \partial _{\xi} \hat \rho _{\xi} (dt ) | \psi ^{\perp}  _{j}  \rangle |^{2} }{\frac{\gamma dt ^{2}}{2}  \! (p _{\ell} + p _{j} ) } 
\nonumber \\
= & \frac{\gamma }{2}  \sum _{\ell =1}^{N} \frac{|\sum _{a,b =1} ^{N} \left [ \Delta C (\xi)    \right ] _{ab}  \langle \psi ^{\perp}  _{\ell}  | \hat Z' _b \left | \phi _{i} \rangle \langle  \phi _{i} \right | \hat Z ' _a | \psi ^{\perp}  _{\ell}  \rangle |^{2}}{p _{\ell} }
. 
\end{align}
Substituting Eq.~\eqref{seq:DeltaC.tog} into the above expression, we compute the analytical form of the short-time-limit of the time-averaged QFI:
\begin{align}
\label{seq:qfipertime.opt.deph.gen}
& \lim _{t\to 0 ^{+}} \left ( \overline{ F _{Q}  } (\hat \rho  _{i}, \mathcal{L} _{\xi};\xi,t)  \right )
\nonumber \\
= & 2 \gamma 
\sum _{j j'} 
\left ( [\partial _{\xi}  g (\xi)]  [\partial _{\xi}  g ^{\dag} (\xi)] \right ) _{j j'}   \text{Tr} (\hat Z'_{j'}  \hat Z'_{j}  \hat\rho _{i}  )
.
\end{align}
For notational simplicity, we denote the above time-averaged QFI as a function $\overline{\mathcal{F}} _{\text{max}} (\hat \rho _{i}, \mathcal{L} _{\xi}  )$ of the initial state $\hat \rho _{i}$ and the Lindbladian $\mathcal{L} _{\xi} $, as
\begin{align}
\label{seq:qfi.avg.rhoi.max.def}
& \overline{\mathcal{F}} _{\text{max}} (\hat \rho _{i}, \mathcal{L} _{\xi}  )
\nonumber \\
\equiv & 2 \gamma 
\sum _{j j'} 
\left ( [\partial _{\xi}  g (\xi)]  [\partial _{\xi}  g ^{\dag} (\xi)] \right ) _{j j'}   \text{Tr} (\hat Z'_{j'}  \hat Z'_{j}  \hat\rho _{i}  )
.
\end{align}
While Ref.~\cite{Zhuang2024} also computed Eq.~\eqref{seq:qfipertime.opt.deph.gen}, that work did not prove optimality of the fast-reset protocol for arbitrary unentangled sensors. In contrast, Ref.~\cite{Gefen2025} proved that the maximal value of Eq.~\eqref{seq:qfi.avg.rhoi.max.def} over all possible initial states $\hat\rho _{i} $ achieves the maximal QFI in Lindbladian channel estimation when we allow arbitrary quantum control, entanglement with ancillae and fast measurements. However, Ref.~\cite{Gefen2025} did not consider cases with a specific initial state; in particular, the approach in Ref.~\cite{Gefen2025} is not applicable to the case when we only allow unentangled sensors.

\subsubsection{Upper bounds on the time-averaged QFI}

In this subsection, we prove the following upper bound on the time-averaged QFI:
\begin{align}
\label{seq:qfi.avg.max}
\overline{ F _{Q}  } (\hat \rho _{i}, \mathcal{L} _{\xi} ;\xi,t ) 
= 
\frac{F _{Q}(\hat \rho _{i}, \mathcal{L} _{\xi} ;\xi,t ) }{t} 
\le & \overline{\mathcal{F}} _{\text{max}} (\hat \rho _{i}, \mathcal{L} _{\xi}  ) 
.
\end{align}
This bound is also tight, as we can achieve equality by a fast-reset protocol [Eq.~\eqref{seq:qfipertime.opt.deph.gen}].

We first express the QFI in terms of the SLD, which will be convenient for the following derivations. 
Recall that the sensor state after evolution for time $t$ reads
\begin{align}
\hat\rho _{\xi } (t) = e ^{\mathcal{L} _{\xi } t} \hat\rho _{i} 
.
\end{align}
To simplify notation, we denote the SLD with respect to $\xi $ as $ \hat S _{\xi } (t) \equiv \mathcal{R} ^{-1} _{\hat \rho _{\xi} (t) } [\partial _{\xi} \hat \rho _{\xi} (t) ]$, so that the QFI in estimating $\xi$ can be written as 
$F _{Q}(\hat \rho _{i}, \mathcal{L} _{\xi} ;\xi,t ) = \text{Tr} [\hat S  _{\xi } (t) ^{2} \hat\rho _{\xi } (t) ]$.
The SLD $ \hat S _{\xi } (t)$ satisfy the following equation:
\begin{align}
\partial _{\xi } \hat\rho _{\xi } (t)= t (\partial _{\xiCorr} \mathcal{L} ) \hat\rho _{\xi } (t)  = \frac{1}{2} \{  \hat S _{\xi } (t),  \hat\rho _{\xi } (t)\}
,
\end{align}
so that we can rewrite the QFI of $\hat\rho _{\xi } (t) $ as~\cite{Lasenby2022}
\begin{align}
\label{seq:qfi.def.SLD}
& F _{Q} (\hat \rho _{i}, \mathcal{L} _{\xi} ;\xi,t ) = \text{Tr} [\hat S  _{\xi } (t) ^{2} \hat\rho _{\xi } (t) ]
\\
= & t \text{Tr} [\hat S  _{\xi } (t) (\partial _{\xiCorr} \mathcal{L} )  \hat\rho _{\xi } (t) ]
.
\end{align}

We first substitute the superoperator $\partial _{\xi} \mathcal{L}  $~\eqref{seq:diffL.def} and $ \Delta C (\xi)$~\eqref{seq:DeltaC.tog} into the QFI in Eq.~\eqref{seq:qfi.def.SLD}, so that we can rewrite the time-averaged QFI as
\begin{align}
& \overline{ F _{Q}  } (\hat \rho _{i}, \mathcal{L} _{\xi} ;\xi,t ) = \text{Tr} [\hat S  _{\xi } (t) (\partial _{\xiCorr} \mathcal{L} )  \hat\rho _{\xi } (t) ]
\nonumber \\
= & \frac{\gamma}{2} \sum _{j\ell} \text{Re}\{ (\Delta C) _{j\ell} 
\text{Tr}  (  [ \hat Z_j ,\hat S  _{\xi } (t) ] \hat Z_\ell  \hat\rho _{\xi } (t)   ) \}
\\
= & \frac{\gamma}{2} \sum _{j\ell} \text{Re}\{ ([\partial _{\xi} g ^{\dag} (\xi) ] g (\xi) + \text{H.c.}) _{j\ell} 
\nonumber \\
& \quad \times \text{Tr}  (  [ \hat Z_j ,\hat S  _{\xi } (t) ] \hat Z_\ell  \hat\rho _{\xi } (t)   ) \}
\\
= & \gamma \sum _{j\ell} \text{Re}\{ ([\partial _{\xi} g ^{\dag} (\xi) ] g (\xi) ) _{j\ell} \text{Tr}  (  [ \hat Z_j ,\hat S  _{\xi } (t) ] \hat Z_\ell  \hat\rho _{\xi } (t)   ) \}
.
\end{align}
The last line is derived using the invariance of trace under cyclic permutations. 
By the Cauchy-Schwarz inequality, we can upper bound the sum in the last line of the equation above as (we omit $\xi$ dependence in $\partial _{\xi} g ^{\dag} $ and $g (\xi)$ matrices and the SLD operator $\hat S  _{\xi } (t) $ in what follows when this does not cause confusion)
\begin{align}
& \sum _{j\ell} \text{Re}\{ ([\partial _{\xi} g ^{\dag} (\xi) ] g (\xi) ) _{j\ell} \text{Tr}  (  [ \hat Z_j ,\hat S  (t) ] \hat Z_\ell  \hat\rho _{\xi } (t)   ) \}
\nonumber \\
\le & \sqrt{\sum _{j\ell} ( g ^{\dag}  g   )_{j\ell}   
\text{Tr}   ( [\hat Z_j  ,\hat S  (t) ] [\hat S  (t)  ,\hat Z_\ell  ] \hat\rho _{\xi } (t) 
 )  }
\nonumber \\
& \times \sqrt{\sum _{ab} ( \partial _{\xi} g ^{\dag}  \partial _{\xi} g   )_{ab}   
\text{Tr}  [\hat Z_a \hat Z_ b \hat\rho _{\xi } (t) ]  }
.
\end{align}
Noting that $\text{Tr}  [\hat Z_a \hat Z_ b \hat\rho _{\xi } (t) ] = \text{Tr}  [\hat Z_a \hat Z_ b \hat\rho _{i} ] $ are invariants of the Lindbladian dynamics, we thus obtain the following bound:
\begin{align}
& \sum _{j\ell} ( g ^{\dag}  g   )_{j\ell}   
\text{Tr}   ( [\hat Z_j  ,\hat S   (t) ] [\hat S  (t)  ,\hat Z_\ell  ] \hat\rho _{\xi } (t) 
 )  
\nonumber \\
\ge & \frac{\left (\frac{\overline{ F _{Q}  } (\hat \rho _{i}, \mathcal{L} _{\xi} ;\xi,t )}{\gamma}   \right ) ^{2}}{\sum _{ab} ( \partial _{\xi} g ^{\dag}  \partial _{\xi} g   )_{ab}   
\text{Tr}  [\hat Z_a \hat Z_ b \hat\rho _{i}  ] }
\\
\label{seq:qfi.prod.bound}
= & \frac{2}{\gamma } \frac{(\overline{ F _{Q}  } (\hat \rho _{i}, \mathcal{L} _{\xi} ;\xi,t )) ^{2}}{\overline{\mathcal{F}} _{\text{max}} (\hat \rho _{i}, \mathcal{L} _{\xi}  )}
.
\end{align}

Using a technique introduced in~\cite{Lasenby2022}, we can relate the time derivative of the QFI to the SLD and the Lindbladian superoperators as 
\begin{align}
& \partial _{t} F _{Q} 
\nonumber \\
=  & \text{Tr} [(\partial _{t} \hat S (t)) \hat S (t) \hat\rho _{\xi } (t) + \hat S (t)  (\partial _{t} \hat S (t)) \hat\rho _{\xi } (t)  + \hat S (t) ^{2} \partial _{t}\hat\rho _{\xi } (t) ]
\nonumber \\
=  & \text{Tr} [ \hat S (t) \partial _{t}(\{ \hat S (t), \hat\rho _{\xi } (t)\} ) - \hat S (t) ^{2}  \mathcal{L} _{\xi } \hat\rho _{\xi } (t)   ]
,
\end{align}
so that we have 
\begin{align}
\partial _{t} F _{Q} = & 2 \text{Tr} [  \hat S (t) (\partial _{\xiCorr} \mathcal{L} )  \hat\rho _{\xi } (t)  + t \hat S (t) (\partial _{\xiCorr} \mathcal{L} )  (\partial _{t} \hat\rho _{\xi } (t) ) ]
\nonumber \\
& - \text{Tr} [\hat S (t) ^{2}  \mathcal{L} _{\xi } \hat\rho _{\xi } (t)   ]
.
\end{align}
Substituting the Lindbladian $\mathcal{L} _{\xi } $ into the last term in the sum above, we obtain
\begin{align}
& \text{Tr} [ \hat S (t) ^{2}  \mathcal{L} _{\xi } \hat\rho _{\xi } (t)  ]
= \text{Tr} [  \hat\rho _{\xi } (t)  \mathcal{L} _{\xi } ^{\dag} (\hat S (t) ^{2} ) ]
\\
= & \text{Tr} [  \{\hat\rho _{\xi } (t) , \hat S (t) \}
\mathcal{L} _{\xi } ^{\dag} (\hat S (t) )  ] 
\nonumber \\
+ & \frac{\gamma}{2} \sum _{j,\ell =1} ^{n} C _{j\ell}  (\xi) 
\text{Tr} \left ( [\hat Z_j ,\hat S (t) ] [\hat S (t)  ,\hat Z_\ell ] \hat\rho _{\xi } (t) 
\right )
\\
= & 2 \text{Tr} [ \hat S (t) \mathcal{L} _{\xi } (\partial _{\xi } \hat\rho _{\xi } (t) ) ]
\nonumber \\
+ & \frac{\gamma}{2} \sum _{j,\ell =1} ^{n} C _{j\ell}  (\xi) 
\text{Tr} \left ( [\hat Z_j ,\hat S (t) ] [\hat S (t)  ,\hat Z_\ell ] \hat\rho _{\xi } (t) 
\right )
.
\end{align}
Noting that 
\begin{align}
& \text{Tr} [  t \hat S (t) (\partial _{\xiCorr} \mathcal{L} )  (\partial _{t} \hat\rho _{\xi } (t) ) ]
= \text{Tr} [  t \hat S (t) (\partial _{\xiCorr} \mathcal{L} )  (\mathcal{L}  _{\xi} \hat\rho _{\xi } (t) ) ]
\nonumber \\
= & \text{Tr} [ \hat S (t) \mathcal{L} _{\xi } (\partial _{\xi } \hat\rho _{\xi } (t) ) ]
,
\end{align}
we have
\begin{align}
& \partial _{t} F _{Q} 
= 2 \text{Tr} [ \hat S (t) \mathcal{L} _{\xi } (\partial _{\xi } \hat\rho _{\xi } (t) ) ]
\nonumber \\
& - \frac{\gamma}{2} \sum _{j,\ell =1} ^{n} C _{j\ell}  (\xi) 
\text{Tr} \left ( [\hat Z_j ,\hat S (t) ] [\hat S (t)  ,\hat Z_\ell ] \hat\rho _{\xi } (t) 
\right )
.
\end{align}

We have thus proved the following relation between the time derivative of the QFI and the time-averaged QFI:
\begin{align}
& \partial _{t} F _{Q} - 2 \frac{F _{Q} }{t}
= t \partial _{t} \left ( \frac{F _{Q} }{t} \right ) - \frac{F _{Q} }{t}
\nonumber \\
= & - \frac{\gamma}{2} \sum _{j,\ell =1} ^{n} C _{j\ell}  (\xi) 
\text{Tr} \left ( [\hat Z_j ,\hat S (t) ] [\hat S (t)  ,\hat Z_\ell ] \hat\rho _{\xi } (t) 
\right )
.
\end{align}
Substituting Eq.~\eqref{seq:qfi.prod.bound} into the equation above, we obtain
\begin{align}
& t \partial _{t} \left ( \frac{F _{Q} }{t} \right ) - \frac{F _{Q} }{t} 
\nonumber \\
\le & - \frac{(\overline{ F _{Q}  } (\hat \rho _{i}, \mathcal{L} _{\xi} ;\xi,t )) ^{2}}{\overline{\mathcal{F}} _{\text{max}} (\hat \rho _{i}, \mathcal{L} _{\xi}  )}
.
\end{align}
For simplicity, we denote $\overline{ F _{Q}  } (\hat \rho _{i}, \mathcal{L} _{\xi} ;\xi,t ) = F _{Q} (\hat \rho _{i}, \mathcal{L} _{\xi} ;\xi,t ) /t $ by $\overline{ F _{Q}  }$ in the following derivations when this does not cause confusion. We have thus shown that
\begin{align}
t\partial _{t} \overline{ F _{Q}  } \le & \overline{ F _{Q}  }- \frac{\overline{ F _{Q}  } ^{2}}{\overline{\mathcal{F}} _{\text{max}} (\hat \rho _{i}, \mathcal{L} _{\xi}  )}
\nonumber \\
=  &  \frac{\overline{ F _{Q}  } ( \overline{\mathcal{F}} _{\text{max}} (\hat \rho _{i}, \mathcal{L} _{\xi}  ) - \overline{ F _{Q}  }) }{\overline{\mathcal{F}} _{\text{max}} (\hat \rho _{i}, \mathcal{L} _{\xi}  ) }
.
\end{align}

The maximal value of the time-averaged QFI $\overline{ F _{Q}  }$ is reached either at an extremal point with $\partial _{t}  (\overline{ F _{Q}  }) =0$, or at a boundary value $t=0$ or $t=\infty$. At extremal time point(s) $\tau_{\text{ext}}$ of the time-averaged QFI, we have 
\begin{align}
& 0= t 
\left .\partial _{t} \overline{ F _{Q}  } \right |_{t=\tau_{\text{ext}}}
\\
\le & \frac{1}{\overline{\mathcal{F}} _{\text{max}} (\hat \rho _{i}, \mathcal{L} _{\xi}  ) }
\left .  \left ( \overline{ F _{Q}  } ( \overline{\mathcal{F}} _{\text{max}} (\hat \rho _{i}, \mathcal{L} _{\xi}  ) - \overline{ F _{Q}  }) \right ) \right |_{t=\tau_{\text{ext}}}
, \nonumber \\
& \Rightarrow \left .\overline{ F _{Q}  }  \right |_{t=\tau_{\text{ext}}}
\le \overline{\mathcal{F}} _{\text{max}} (\hat \rho _{i}, \mathcal{L} _{\xi}  ) 
.
\end{align}
From Eq.~\eqref{seq:qfipertime.opt.deph.gen}, we can explicitly compute the time-averaged QFI in the $t\to 0^{+}$ limit as
\begin{align}
\lim _{t\to 0^{+}} \overline{ F _{Q}  }  
=  \overline{\mathcal{F}} _{\text{max}} (\hat \rho _{i}, \mathcal{L} _{\xi}  )  
.
\end{align}
Combining above results with the fact that the time-averaged QFI decays exponentially to zero in the long time ($t \to \infty$) limit, we can bound the time-averaged QFI with arbitrary evolution times as
\begin{align}
\overline{ F _{Q}  }  (\hat \rho _{i}, \mathcal{L} _{\xi} ;\xi,t ) 
\le & \overline{\mathcal{F}} _{\text{max}} (\hat \rho _{i}, \mathcal{L} _{\xi}  )
,
\end{align}
which proves Eq.~\eqref{seq:qfi.avg.max}.

\subsubsection{Case study: $2$-qubit Lindbladian $\mathcal{L} _{\xiCorr,\text{2qb}}$ and multiqubit Lindbladian $\mathcal{L} _{\xiCorr,N\text{qb}}$}

\label{sisec:2qb.epsilon.gen}

We can use Eq.~\eqref{seq:qfipertime.opt.deph.gen} to calculate the optimal QFI when constraining the total protocol time. For the two-qubit $\mathcal{L} _{\xiCorr,\text{2qb}} = \mathcal{L} _{\xiCorr} [C _{\text{2qb}} (\xiCorr)  ] $ discussed in the main text, where $[C _{\text{2qb}} (\xiCorr) ]_{j\ell} = 1 - \xiCorr (1-\delta _{j\ell}) $, we have 
\begin{align}
& \max _{ t} \overline{ F _{Q}  } 
( |\phi _{i} \rangle  , \mathcal{L} _{\xiCorr,\text{2qb}} ;\xiCorr,t) \\
= & 2 \gamma 
\sum _{j j'} 
\left ( [\partial _{\xi}  g (\xi)]  [\partial _{\xi}  g ^{\dag} (\xi)] \right ) _{j j'} \langle \phi _{i} | \hat Z'_{j'}  \hat Z'_{j}  |\phi _{i} \rangle
\\
= &  \frac{\gamma}{4} \left ( \frac{\langle \phi _{i} | ( \hat Z'_{1} + \hat Z'_{2} )^2 |\phi _{i} \rangle}{2- \xi} + \frac{\langle \phi _{i} | ( \hat Z'_{1} - \hat Z'_{2} )^2 |\phi _{i} \rangle}{\xi}\right )
,
\end{align}
where the shifted jump operators' expectation values satisfy $\langle \phi _{i} | \hat Z'_{j'}  \hat Z'_{j}  |\phi _{i} \rangle = \langle \phi _{i} | \hat Z_{j'}  \hat Z_{j}  |\phi _{i} \rangle - \langle \phi _{i} | \hat Z_{j'}  |\phi _{i} \rangle \langle \phi _{i} |  \hat Z_{j}  |\phi _{i} \rangle$. 
The optimal time-averaged QFIs over all states and for all separable states can be computed,  respectively, as 
\begin{align}
\max _{|\phi _{i} \rangle , t} &\overline{ F _{Q}  } 
( |\phi _{i} \rangle  , \mathcal{L} _{\xiCorr,\text{2qb}} ;\xiCorr,t) =\frac{\gamma}{{\xiCorr  }}
,\\
\max _{| \phi_{1} \rangle , |\phi_2 \rangle , t} & \overline{ F _{Q}  } 
( | \phi_{1} \rangle \! \otimes \! | \phi_{2} \rangle , \mathcal{L} _{\xiCorr,\text{2qb}} ;\xiCorr,t) =\frac{\gamma}{{\xiCorr (2- \xiCorr)}}
,
\end{align}
which reproduces the results in the main text.

In the case of the multiqubit dephasing Lindbladian $\mathcal{L} _{N\mathrm{qb} } $ discussed in the main text, the coefficient matrix is given by $C _{N\text{qb}} (\xiCorr) =  \mathbb{I} _{N} - \frac{1-\xiCorr}{N}\sum_{j,\ell }
\left | j\rangle \langle \ell\right | $. We thus obtain the optimal QFI when using only separable input states $\hat \rho _{\text{sep}}$ and the optimal QFI over all quantum states as  
\begin{align}
\max \limits_{\hat \rho  _{i} \in \hat \rho _{\text{sep}} ; t} 
\overline{ F _{Q}  } (\hat \rho  _{i}, \mathcal{L} _{\xiCorr,N\text{qb}} ;\xiCorr,t) 
& = \frac{\gamma }{2} \frac{1}{\xiCorr }
, \\ 
\max \limits_{\hat \rho  _{i}; t} \overline{ F _{Q}  } (\hat \rho  _{i}, \mathcal{L} _{\xiCorr,N\text{qb}} ;\xiCorr,t) 
& = \frac{\gamma }{2}  \frac{N}{\xiCorr }
. 
\end{align}
We see that, in this case, using an entangled state leads to a factor-$N$ enhancement in the QFI, which is the same scaling as the standard Heisenberg limit, although the QFI does not have Heisenberg scaling with respect to total measurement time. In the time-limited regime, entanglement advantage in estimating correlated dephasing in Eq.~\eqref{seq:diffL.def} can have at most a Heisenberg-type scaling [i.e., $O(N)$], whereas we can achieve exponential entanglement advantage in the shot-limited regime.

\begingroup
\renewcommand{\thesubsection}{III}
\subsection{Generalizations and robustness of exponential entanglement advantage in sensing noise in the center-of-mass-like mode in the presence of
a many-body Lindbladian gap}
\label{subsec:corr.deph.mat.gen}
\endgroup

Here, we prove that the exponential entanglement advantage discussed in the main text can be generalized to a broader class of correlated dephasing Lindbladians of the form $\mathcal{L} _{N\mathrm{qb} } = \mathcal{L} _{\xi} [C _{N\text{qb}} (\xiCorr)]$ in Eq.~\eqref{seq:multi.dephL.def}, as long as dephasing rates generated from such a Lindbladian have a large many-body gap and the state with lowest dephasing rate corresponds to a multipartite entangled state. We also show that the exponential entanglement advantage holds in the absence of prior knowledge about the structure of the noise correlation matrix $C(\xi)$.

We start by discussing the microscopic origin of such correlated noise environments. As shown in Sec.~I, for sensor qubits with an appreciable energy separation, the dominant source of decoherence, as seen by the qubits, is dephasing noise, where the dephasing coefficient matrix elements are given by the low-frequency noise spectrum of the environment (i.e., the sensing target):
\begin{align}
\frac{\gamma}{2} [C _{N\text{qb}} (\xiCorr) ]_{j\ell} = 
\gamma ^{(z)}_{j\ell} & = \lim \limits _{\omega \to 0} \overline{\mathcal{S}} ^{(zz)}_{j\ell} [\omega] 
.
\end{align}
For convenience of analysis, we assume the sensing target has translational invariance; however, the discussion below can be directly extended to target systems without exact spatial translational symmetry, as long as the spatial structure of the quasiparticle excitations in the system is known. Specifically in the case with spatial translational invariance, for the purpose of probing two-point correlation functions of the noise operators, it suffices to place the sensors periodically in real space, so that we can diagonalize the dephasing coefficient matrix in terms of the momentum-basis vectors $\mathbf{e} _{k} = ({ 1 }/{\sqrt{N}})  \sum  _{j =1} ^{N} e ^{i\frac{2 \pi k}{N}j} \mathbf{e} _{j} $ as (we omit $\xiCorr$ dependence when this does not cause confusion)
\begin{align}
\frac{\gamma}{2} C _{N\text{qb}}
= \sum _{k} \gamma _{k} \mathbf{e} _{k} \mathbf{e} _{k} ^{\dag}
. 
\end{align}
The eigenvalues $\gamma _{k} $ now correspond to the noise spectral densities of the momentum-space noise operators 
$\hat B  _{z,k} = ({ 1 }/{\sqrt{N}})  \sum  _{j =1} ^{N} e ^{i\frac{2 \pi k}{N}j} \hat B  _{z,j} $, as (cf.~Eq.~\eqref{seq:nsd.gen.def})
\begin{align}
\label{seq:Cti.gammak}
& \gamma _{k}  = \lim \limits _{\omega \to 0} \overline{\mathcal{S}} ^{(zz)}_{kk } [\omega] 
\nonumber \\ = & 
\frac{1}{2} 
\lim \limits _{\omega \to 0} \int ^{\infty}_{ - \infty  } 
\langle  \{  \hat B _{z,k} (s) ,
\hat B  _{z,k} (0) 
\}  \rangle e^{i \omega s } ds  
. 
\end{align}
We now consider the scenario of probing the noise induced by an interacting many-body (classical or quantum) system as the sensing target. Measuring the eigenvalues $\gamma _{k} $ of the noise coefficient matrix  then translates to measuring the number of the quasiparticle excitations in momentum space of the sensing target, which can be used to measure, e.g., the temperature of the system. Furthermore, with the aid of dynamical decoupling techniques~\cite{Hirayama2011,Suter2011}, such noise measurements can be used to reconstruct the noise spectral density function $\overline{\mathcal{S}} ^{(zz)}_{kk } [\omega] $ at finite frequencies~\cite{Oliver2011}, which can be utilized to detect whether the system is in thermal equilibrium.

In general, the reconstruction of $\gamma _{k} $ involves a multiparameter estimation problem. In principle, we can measure the dephasing rates of various superposition states $| \Psi _{\boldsymbol{\alpha, \beta}} \rangle = (| \boldsymbol{\alpha } \rangle + | \boldsymbol{\beta} \rangle) / \sqrt{2}$ of the multiqubit sensor and relate the measured dephasing rates to $\gamma _{k} $ via the formula $\gamma _{\boldsymbol{\alpha} ,\boldsymbol{\beta}} =  \frac{\gamma }{4}
(\boldsymbol{\alpha} - \boldsymbol{\beta} ) ^{T} C _{N\text{qb}}  (\boldsymbol{\alpha} - \boldsymbol{\beta} ) $. Recall that $\gamma _{k} $ are eigenvalues of the noise correlation matrix (see Eq.~\eqref{seq:Cti.gammak}). We thus have
\begin{align}
\gamma _{\boldsymbol{\alpha} ,\boldsymbol{\beta}} = \frac{1 }{2} \sum _{k} \gamma _{k}   |\mathbf{e} _{k} ^{T} (\boldsymbol{\alpha} - \boldsymbol{\beta} )| ^{2}
.
\end{align}
If the values of $\gamma _{k} $ are all comparable, this procedure can be done efficiently. However, if one of the eigenvalues $\gamma _{k _{0}} $ is much smaller than all the other ones, then its contribution to the dephasing dynamics of a generic superposition state $| \Psi _{\boldsymbol{\alpha, \beta}} \rangle$ will be greatly suppressed by the other eigenmodes, unless the state dephasing rate solely depends on $\gamma _{k _{0}} $ and is independent of the rest of the noise modes ($k \ne k _{0}$). Whenever such a gap exists between the slowest dephasing rate, corresponding to coherence component $\ketbra{\boldsymbol{\alpha } _{k _{0}}}{\boldsymbol{\beta }_{k _{0}}}$, and the dephasing of other coherences, we can apply an analysis similar to the one presented in the main text to prove that the optimal shot-limited sensing strategy is given by initial sensor state $| \Psi _{\boldsymbol{\alpha _{k _{0}}, \beta _{k _{0}}}} \rangle$ evolving for a time $t$ that maximizes the quantum Fisher information of state $ e ^{ \mathcal{L} _{N\mathrm{qb} } t } (| \Psi _{\boldsymbol{\alpha _{k _{0}}, \beta _{k _{0}}}} \rangle)$ with respect to $\gamma _{k _{0}} $. Thus, we recover a single-parameter estimation task with the target parameter $\xiCorr = \gamma _{k _{0}} $. The entanglement advantage in this case can be computed as 
\begin{align}
& \lim _{\xiCorr \to 0}
\frac{\max \limits_{\hat \rho  _{i}, t} F _{Q}  (\hat \rho  _{i}, \mathcal{L} _{\xiCorr,N\text{qb}} ;\xiCorr,t) }
{\max \limits_{\hat \rho  _{\text{sep}}, t} F _{Q}  
( \hat \rho  _{\text{sep}} , \mathcal{L} _{\xiCorr,N\text{qb}} ;\xiCorr,t) }
\nonumber \\
= & \left (  \max _{\hat \rho  _{\text{sep}}}  \left \langle  \Psi _{\boldsymbol{\alpha _{k _{0}}, \beta _{k _{0}}}} | \hat \rho  _{\text{sep}}| \Psi _{\boldsymbol{\alpha _{k _{0}}, \beta _{k _{0}}}} \right  \rangle \right ) ^{-1}
.
\end{align}
When the superposition state $ | \Psi _{\boldsymbol{\alpha _{k _{0}}, \beta _{k _{0}}}} \rangle$ corresponding to the lowest dephasing rate is an entangled state, we expect a nontrivial entanglement advantage. For $ | \Psi _{\boldsymbol{\alpha _{k _{0}}, \beta _{k _{0}}}} \rangle$ exhibiting multipartite entanglement that scales with the number of sensor qubits, we further expect a scalable entanglement advantage.

As discussed in the main text, one specific class of Lindbladians that hosts a finite gap between the smallest dephasing rate (or equivalently, the smallest nonzero Lindbladian eigenvalue) and the second smallest dephasing rate is the following translationally invariant correlated dephasing dynamics [see Eq.~\eqref{eq:lindblad.nqb.def} in the main text]
\begin{align}
\frac{2}{\gamma} \mathcal{L} _{N\mathrm{qb} } & = 
\xiCorr \mathcal{D} [\hat Z_{k=0} ]  + \sum _{k \ne 0} \zeta _{k} \mathcal{D} [\hat Z_k ] .
\end{align}
The noise correlation matrix reads $C _{N\text{qb}}  = \frac{\gamma }{2} (\xiCorr \mathbf{e} _{k=0} \mathbf{e} _{k=0} ^{\dag} + \sum _{k \ne 0 } \zeta _{l} \mathbf{e} _{k} \mathbf{e} _{k} ^{\dag}) $, with $\mathbf{e} _{k} = \frac{1}{{\sqrt{N}}}  \sum  _{j =1} ^{N} e ^{i\frac{2 \pi k}{N}j} \mathbf{e} _{j} $. Thus, we obtain an exponential entanglement advantage whenever $\zeta _{k} \gg \xiCorr$, $\forall k \ne 0 $. A corollary of this result is that the exponential entanglement advantage is robust against a finite value of uncorrelated dephasing and amplitude damping, up to a threshold dissipation rate $O( \gamma \xi) $. Similarly, for weak additional amplitude-damping, its leading-order impact on the final sensor state's sensitivity with respect to $\xiCorr$ enters through an extra relaxation-induced dephasing rate, so the same gap-based argument implies robustness up to a threshold dissipation rate $O( \gamma \xi) $.

\begin{figure}[t]
    \centering
    \includegraphics[width=\columnwidth]{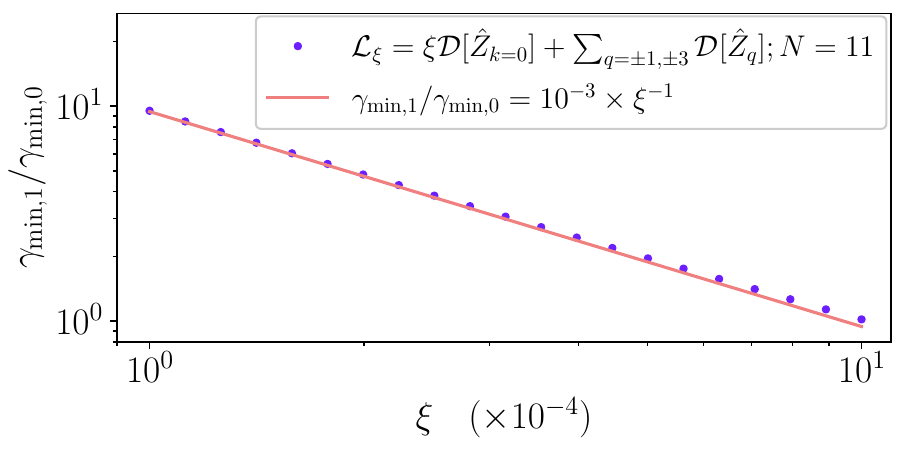}
    \caption{The ratio of the second smallest nonzero dephasing rate ($\gamma _{\text{min},1} $) and the smallest nonzero dephasing rate ($\gamma _{\text{min},0} $) of the translationally invariant correlated dephasing Lindbladian $\xiCorr \mathcal{D} [\hat Z_{k=0} ]  + \sum _{k =\pm 1, \pm3} \mathcal{D} [\hat Z_k ] $, with $N=11$. Purple dots show numerical results from exactly diagonalizing the Lindbladians, whereas the light red line depicts an inverse linear fit function of the form $\gamma _{\text{min},1}/\gamma _{\text{min},0}  \propto \xi ^{-1}$. As $\xiCorr$ decreases, the ratio $\gamma _{\text{min},1}/\gamma _{\text{min},0} $ increases approximately as $\xi ^{-1}$, indicating the presence of a robust gap in the many-body dephasing rates. Moreover, in this Lindbladian, the smallest dephasing rate always corresponds to the decoherence of an $N$-qubit GHZ state. These features ensure an exponential entanglement advantage in estimating $\xiCorr$ in the shot-limited regime (see Sec.~\ref{subsec:corr.deph.mat.gen} for details). 
    }
    \label{suppfig:lind.gap.N11}
\end{figure}

It is natural to ask whether this exponential entanglement advantage survives if the noise correlation matrix is no longer near-maximally correlated, e.g., when not all the  $k \neq 0$ jump operators are present. Finite-size numerical evidence with up to $N=13$ qubit sensors suggests that when $N$ is odd, Lindbladians of the form $\xiCorr \mathcal{D} [\hat Z_{k=0} ]  + \sum _{k =\pm 1, \pm3} \mathcal{D} [\hat Z_k ] $ with only two pairs of $k \ne 0$ jump operators also enable the exponential entanglement advantage, see Fig.~\ref{suppfig:lind.gap.N11}. Alternatively, exponential entanglement advantage can also survive if the noise correlation matrix has an additional power-law decay envelope $C_{j\ell} \sim |j-\ell|^{-\alpha} $, as long as the exponent $\alpha$ is sufficiently small. However, rigorously verifying those observations for larger system sizes is numerically challenging, as it requires solving a discrete optimization problem (i.e., finding the two lowest eigenvalues of the Lindbladian, which acts on an exponentially large space). It would be interesting to explore these questions further in future work.

Lastly, we remark that our results can be straightforwardly generalized to the case of multiparameter estimation. More specifically, the single-parameter estimation results in our work assume prior knowledge about all other parameters in the problem. Without such knowledge, the problem becomes a genuine multiparameter estimation task, which may introduce an extra poly($N$) time (or shot) resource overhead. However, such overhead contributes only a subleading correction if there is an exponential entanglement advantage in estimating any of the given parameters. As such, the exponential separation between entangled and unentangled sensors' performance holds regardless of the prior knowledge of the other $C _{j\ell} $ matrix elements (or similarly, the fluctuations in the nonzero momentum modes $\hat B _{k\ne 0}$), since these terms could be estimated using product-state dephasing measurements with at most polynomial overhead.
\\

\begingroup
\renewcommand{\thesubsection}{IV}
\subsection{Entanglement advantage in estimating bosonic loss, gain, and random displacement Lindbladians}
\endgroup

Here, we discuss entanglement enhancement in estimating bosonic Gaussian Lindbladians including correlated bosonic loss, gain, and random displacements, which are distinct from the pure-dephasing noise discussed in the main text. As we show, the optimal sensing protocol differs between the time-limited and the shot-limited regimes, similarly to the pure-dephasing case. However, the entanglement advantage now scales at most linearly in the number of modes participating in the dissipative dynamics, as opposed to an exponential separation.

We consider an $N$-mode bosonic sensor system with lowering operators $\hat a _{\ell}$. The target noisy dynamics follows time evolution generated by the Lindbladian
\begin{align}
\label{seq:qme.boson.def}
\mathcal{L} _{\xi} [C(\xi)]  \hat\rho  = \! 
\frac{\gamma}{2}  \! \sum _{j,\ell =1} ^{N} C _{j\ell}  (\xi) \left( \hat o _\ell \hat\rho \hat o_j ^{\dag} -\frac{\{ \hat o_j ^{\dag}  \hat o _\ell , \hat \rho\}}{2} \right) 
,
\end{align}
where $\hat o_j $ denote bosonic lowering ($\hat o_j = \hat a_j $), raising ($\hat o_j = \hat a_j ^{\dag} $) and quadrature operators ($\hat o_j = c _j \hat a_j + \text{H.c.} $) on the $j$-th mode for loss, gain, and random displacement dynamics, respectively, and we assume $C(\xi) = C(0)+ \xi \Delta C$. Our goal is to compute the quantum Fisher information (QFI) with respect to parameter $\xi$, 
\begin{align}
F _{Q} 
(\hat \rho _{i}, \mathcal{L} _{\xi} ;\xi,t )   
\equiv & 8 \lim _{d \xi \to 0}
\frac{ 1- \sqrt{ F (e ^{\mathcal{L} _{\xi} t} \hat \rho  _{i}, e ^{\mathcal{L} _{\xi + d \xi } t} \hat \rho  _{i} ) } }{ d \xi ^{2}} 
,
\end{align}
which sets the fundamental limit on sensitivity in the shot-limited regime, as well as the time-averaged QFI, $\overline{ F _{Q}  } (\hat \rho _{i}, \mathcal{L} _{\xi} ;\xi,t )  = F _{Q} 
(\hat \rho _{i}, \mathcal{L} _{\xi} ;\xi,t )  /t $, in the time-limited regime (see also Eq.~\eqref{neq:chnl.qfi.def} in the main text).

For the purpose of discussions in this section, we focus on Gaussian initial states, which can be efficiently prepared via linear optical operations. Such states can be fully characterized by the means $\langle \hat x _{\ell} \rangle $, $\langle \hat p _{\ell} \rangle $ and variances $\langle \{ \delta \hat m _{\alpha,\ell}, \delta \hat m _{\beta, j} \} \rangle $ ($\delta \hat m _{\alpha, \ell} = \hat x _{\ell}$, $\delta \hat m _{\beta, j} = \hat p _{j}$) of the bosonic quadrature operators, $\hat x _{\ell}$ and $\hat p _{\ell}$, in the initial state: 
\begin{align}
& \hat x _{\ell} = \frac{\hat a _{\ell}+ \hat a ^{\dag} _{\ell}}{ \sqrt{2}}
, \quad 
\hat p _{\ell} = i \frac{\hat a ^{\dag} _{\ell} - \hat a _{\ell}}{ \sqrt{2}}
, \\
& \delta \hat m _{\beta, j} = \hat m _{\beta, j} - \langle \hat m _{\beta, j}\rangle 
. 
\end{align}
We focus on the scenario where $[C(0), \Delta C]=0$, so that the two matrices can be diagonalized in a mutual orthonormal basis $\{\boldsymbol{\eta} _{\ell} \}$ simultaneously as
\begin{align} 
C(0)  = \sum _{\ell }
\epsilon _{\ell} \boldsymbol{\eta} _{\ell} \boldsymbol{\eta} _{\ell} ^{\dag}
, \quad 
\Delta C = \sum _{\ell }  
\upsilon _{\ell} \boldsymbol{\eta} _{\ell} \boldsymbol{\eta} _{\ell} ^{\dag}
,
\end{align}
Making use of the results in Appendix D of Ref.~\cite{Lasenby2022}, the time-limited QFI is upper bounded by 
\begin{align}
\label{seq:qfi.gaus.ineq}
\partial _{t} F _{Q} 
(\hat \rho _{i}, \mathcal{L} _{\xi} ;\xi,t ) & \le 
\sum _{\ell } \frac{\upsilon _{\ell} ^{2}}{\epsilon _{\ell}}
\text{Tr} (\hat L _{\ell} ^{\dag} \hat L _{\ell} e ^{t \mathcal{L} _{\xi} } \hat \rho _{i} )
, \\
\hat L _{\ell} &  = \sum _{b} \mathbf{e} _{b }^{\dag} \boldsymbol{\eta} _{\ell} \hat o _{b}
.
\end{align}

We can use the upper bound in Eq.~\eqref{seq:qfi.gaus.ineq} to deduce optimal estimation strategies in the time-limited regime. For bosonic loss dynamics with $\hat o _{\ell} = \hat a _{\ell} $, the time derivative of QFI in Eq.~\eqref{seq:qfi.gaus.ineq} is monotonically decreasing in time, so that the optimal time-averaged QFI satisfies the inequality
\begin{align}
\overline{ F _{Q}  } (\hat \rho _{i}, \mathcal{L} _{\xi} ;\xi,t )
\le \sum _{\ell } \frac{\upsilon _{\ell} ^{2}}{\epsilon _{\ell}} \text{Tr} (\hat L _{\ell} ^{\dag} \hat L _{\ell}  \hat \rho _{i} )
,
\end{align}
with the upper bound saturated by a repetitive reset protocol. In particular, the time-average QFI averaged over the total boson number is given by 
\begin{align}
\frac{\overline{ F _{Q}  } (\hat \rho _{i}, \mathcal{L} _{\xi} ;\xi,t )}{\sum _{\ell } \text{Tr} (\hat a _{\ell} ^{\dag} \hat a _{\ell}  \hat \rho _{i} )}
\le \max _{\ell } \frac{\upsilon _{\ell} ^{2}}{\epsilon _{\ell}} 
,
\end{align}
and we can achieve this optimal time-averaged QFI per boson by concentrating all the excitations into the collective mode that maximizes the ${\upsilon _{\ell} ^{2}}/ {\epsilon _{\ell}}$ ratio.

Entanglement advantage in the time-limited regime for bosonic displacement Lindbladians (i.e., the jump operators are now quadrature operators with $\hat o_j = c _j \hat a_j + \text{H.c.}  $, where $c_{j} $ are complex constant coefficients) has been discussed in Ref.~\cite{Zhuang2024}; here, we focus on entanglement advantage for pure-loss Lindbladians in the shot-limited regime. As an example, we may consider the specific case where $C(0) $ and $\Delta C$ are rank-$1$ matrices, 
$C(0)  = \epsilon _{0} \boldsymbol{\eta} _{0} \boldsymbol{\eta} _{0} ^{\dag}$
, $\Delta C =  
\upsilon _{0} \boldsymbol{\eta} _{0} \boldsymbol{\eta} _{0} ^{\dag}$. Fixing the total boson number, we derive the entanglement advantage ratio in this case to be
\begin{align}
& 
\frac{\left.\max \limits_{\hat \rho  _{i}, t} F _{Q}  (\hat \rho  _{i}, \mathcal{L} _{\xiCorr } ;\xiCorr,t)\right| _{ \sum _{\ell }  
\text{Tr}(\hat a _{\ell} ^{\dag} \hat a _{\ell}  \hat \rho  _{i} )   = \bar{n}_{\text{tot}}} }
{\left. \max \limits_{\hat \rho  _{\text{sep}}, t} F _{Q}  
( \hat \rho  _{\text{sep}} , \mathcal{L} _{\xiCorr } ;\xiCorr,t) \right| _{ \sum _{\ell }  
\text{Tr}(\hat a _{\ell} ^{\dag} \hat a _{\ell} \hat \rho  _{\text{sep}} ) = \bar{n}_{\text{tot}}} }
\nonumber \\
= & \left [ \sum _{a} (\mathbf{e} _{a }^{\dag} \boldsymbol{\eta} _{0}) ^{2} \right ] ^{-1}
.
\end{align}
Thus, we see that the entanglement advantage scales at most linearly with respect to the number of bosonic modes, in contrast to the exponential entanglement advantage for the pure-dephasing qubit noise discussed in the main text. This is because the Lindbladian spectrum of the photon loss dynamics in Eq.~\eqref{seq:qme.boson.def} is continuous and does not exhibit a finite gap. Alternatively, if we take instead $\hat o _{\ell} = \hat n _{\ell} = \hat a _{\ell} ^{\dag} \hat a _{\ell} $, then a similar exponential entanglement advantage again emerges for the correlated noise correlation matrices satisfying the condition discussed in Sec.~\ref{subsec:corr.deph.mat.gen}.

In contrast, the optimal probing strategy for estimating heating dissipative dynamics differs from the loss case. Specifically, the right-hand side of Eq.~\eqref{seq:qfi.gaus.ineq} shows that the QFI growth rate for the pure-heating or random displacement dynamics can increase indefinitely in time. This means that, to maximize QFI, one should choose the evolution time $t$ to be as large as possible in both the time-limited and the shot-limited regimes. Such exotic QFI time dependence is due to the infinite dimensional Hilbert space, and the super-linear growth of QFI in the heating case accompanies an exponential increase in the photon number. Practically, we expect such growth to eventually terminate due to residual excitation loss and/or higher-order nonlinear processes.

\bibliography{qsense_noise}

\end{document}